\documentclass[showpacs,preprintnumbers,amsmath,amssymb]{revtex4}
\usepackage{amsfonts,amsthm}
\begin{document}
\newtheorem{definition}{Definition}
\newtheorem{theorem}{Theorem}
\newtheorem{proposition}{Proposition}
\newtheorem{lemma}{Lemma}
\def\var{\varepsilon}

\title{Dynamical detailed balance and local KMS condition for
non-equilibrium states}
\author{Luigi Accardi}%
\email{accardi@volterra.mat.uniroma2.it}
\author{Kentaro Imafuku}%
 \email{imafuku@volterra.mat.uniroma2.it}
\affiliation{%
Centro Vito Volterra, Universit\'{a} di Roma Torvergata\\
I00133 Roma, Italy
}%

\date{
\today
}
\begin{abstract}
The principle of detailed balance is at the basis of equilibrium physics and
is equivalent to the Kubo-Martin-Schwinger (KMS) condition (under quite general
assumptions). In the present paper we prove that a large class of open quantum
systems satisfies a dynamical generalization of the detailed balance condition
({\it dynamical detailed balance})
expressing the fact that all the micro-currents, associated to the
Bohr frequencies are constant.
The usual (equilibrium) detailed balance condition is
characterized by the property that this constant is identically zero.
From this we deduce a simple and experimentally measurable relation
expressing the microcurrent associated to a transition between
two levels $\epsilon_m\rightarrow\epsilon_n$ as a linear combination of
the occupation probabilities
of the two levels, with coefficients given by the generalized susceptivities
(transport coefficients). Finally, using a master equation
characterization of the dynamical detailed balance condition, we show that this condition is equivalent to a "local" generalization of the usual KMS condition.
\end{abstract}
\maketitle
\section{introduction}
\label{sec:intro}
To understand non-equilibrium phenomena is one of the most important
challenges of modern physics. The monographs
\cite{Prigogine62,Zubarev,Toda} summarize the early developments in
this direction and, after them, several endeavors were
made by many authors to construct a satisfactory description of
non-equilibrium phenomena from the stand point of microscopic physics
(cf. e.g. \cite{Spohn-Lebowits78,Antoniou,Tasaki,Jalsic02,Schmuser02,Bedeaux01,Ojima}). 
As pointed out by many authors (see for example \cite{Ojima}),
the most crucial difficulty of the problem is that we lack a good
characterization of non-equilibrium states whereas we have criteria
for the equilibrium case: detailed balance, the
KMS condition, stability and so on.
In the present paper, starting from some physically
interesting situations we deduce a general characterization
of a class of stationary states
which satisfy a condition (dynamical detailed balance) generalizing
the usual detailed balance and KMS conditions. 
For this purpose, we apply the stochastic limit technique
\cite{stochastic_limit,AcLuVo,Accardi-Fagnola,Accardi-Imafuku-Kozyrev02,Accardi-Imafuku-Lu02} to some concrete and widely studied models
and show that this leads to a natural
generalization of both the detailed balance and the KMS conditions
which characterizes a 
rather wide and interesting class of non-equilibrium stationary states.

The first basic idea of the present paper can be described as follows.
The most commonly used states in quantum field theory are the Fock
(vacuum) or Gibbs (equilibrium) states. When a field in such a state
interacts with a discrete system (e.g. an atom) in the stochastic
limit one obtains a master equation for the system whose
stationary state is
the ground state of the atom, if the field  was originally in the Fock
state; while it is the Gibbs
state of the system at inverse  temperature $\beta$, if
the field was originally in its equilibrium state at inverse temperature
$\beta$. The systematic development of the theory of stochastic limit
(see below and \cite{stochastic_limit,AcLuVo}),
has revealed that the above described phenomenon is quite universal namely: for a large class of states 
(including many concrete examples which are neither Fock nor equilibrium)
the stochastic limit procedure allows to deduce master equations whose
associated Markov semigroups drive the system to a stationary state
$\rho_\infty$ in the sense that, independently of the initial state 
$\rho_0$, one has $$\lim_{t\to+\infty}P^t_{*}\rho_0=\rho_\infty$$
($P^t_*$ is the Markov semigroup acting on density matrices).
This fact suggests to give a dynamical characterization of ground (or
equilibrium) states of the system (atom) in terms of their response to
an interaction with the environment (field) in the stochastic limit
regime. This extends to the non-equilibrium regime the approach of
\cite{Gorini-Kossokowski76,Kossakowski-Frigerio-Gorini-Verri77}.
In fact, from the above considerations it is natural to expect that the
analysis of the stationary states of master equations associated via stochastic
limit to non equilibrium states of the environment, will lead to
a new class of states, of discrete quantum systems, which should play
for non-equilibrium phenomena, a role analogue to that played by Gibbs
states for equilibrium phenomena.
In the present paper, we prove that this is indeed the case.

The second basic idea of the present paper is to exploit 
the main advantage of stochastic limit with respect to the old
Markovian approximation namely: 
the field degrees of freedom are not traced away, but they survive in the
limit as "quantum noise" (or master field). In particular, as shown in
\cite{Accardi-Imafuku-Kozyrev02,Accardi-Imafuku-Lu02}, the slow degrees of
freedom of the field (e.g. the functions of the free energy of the field)
survive in the stochastic limit.
This allows us to define the energy currents in a natural way and to study
their dynamics, thus going far beyond the Markov approximation
where one only obtains the master equation for system observables
and looses any control on the limits of field observables.

We will illustrate our ideas with two models:
one is very well studied in the literature
and consists of a system interacting with two equilibrium thermal
reservoirs at different temperatures. 
The master equation
approach to this model was discussed in \cite{Spohn-Lebowits78}.
As already mentioned,
this technique cannot be applied to the problem studied in the
present paper, i.e. the dynamical study of the currents associated to
the field because the field degrees of freedom are traced away
from the beginning.
The second class of models is more general (see Sec.\ref{ModelB}),
because the field, with which the system interacts
is not in a usual equilibrium state,
but in a new class of states in which, roughly speaking,
{\it each frequency is at local equilibrium at its own (frequency dependent)
temperature}. Although states of this type have been considered in studies
of molecular kinetics, we do not know if these states have been
experimentally realized.
However their structure, characterized by local equilibrium at energy
dependent
temperatures, is a natural modification of the usual equilibrium states
(see Sec.~\ref{sec:sum} below) and we are confident that
the inventiveness of experimentalists is rich enough to allow their
realization.

We briefly describe the general scheme of the stochastic
limit technique for  Hamiltonians of the form
\begin{equation}
H^{(\lambda)}=H_0+\lambda H_I
\end{equation}
where $\lambda$ is a real parameter, $H_0$ is the free Hamiltonian and 
$H_I$ is the interaction Hamiltonian (see the concrete example
in the next section).  The general idea of the stochastic 
limit approach \cite{stochastic_limit,AcLuVo} is to introduce the
time rescaling
\begin{equation}\label{eq:rescaling}
t\rightarrow t/\lambda^2
\end{equation}
in the solution 
\begin{equation}
U^{(\lambda)}_t=e^{itH_0}e^{-it H^{(\lambda)}}
\end{equation}
of the Schr\"{o}dinger equation in interaction picture associated to 
the Hamiltonian $H^{(\lambda)}$, i.e.
\begin{equation}
\frac{d}{dt}U_t^{(\lambda)}=-i\lambda H_I(t) U_t^{(\lambda)},\quad
H_I(t)=e^{itH_0}H_Ie^{-itH_0}.
\end{equation}
The rescaling (\ref{eq:rescaling}) gives the rescaled equation
\begin{equation}
\frac{d}{dt}U^{(\lambda)}_{t/\lambda^2}=-\frac{i}{\lambda}
H_I(t/\lambda^2)U^{(\lambda)}_{t/\lambda^2}
\end{equation}
and the limit $\lambda\rightarrow 0$ (which is equivalent to
$\lambda\to0$ and $t\rightarrow \infty$ under
the condition that $\lambda^2 t$ tends to a constant) captures the dominating
contributions to the dynamics, which, under appropriate assumptions on
the model \cite{AcLuVo} is shown to converge to the solution of
\begin{equation}
\frac{d}{dt}U_t=-i h_t U_t,\quad h_t=\lim_{\lambda\rightarrow 0}
\frac{1}{\lambda}H_I(t/\lambda^2)\ ,\quad U(0)=1\label{7}
\end{equation}
Similarly one obtains the limit of the Heisenberg evolution
\begin{equation}
\lim_{\lambda\to0}X^{(\lambda)}_t :=\lim{U^{(\lambda)}_{t/\lambda^2}}^\dagger X
U^{(\lambda)}_{t/\lambda^2}= U_t^\dagger X U_t\label{6}
\end{equation}
where $U_t$ is the solution of (\ref{7}) and $X$ is an observable
belonging to a certain class (slow observables, cf. Sec.\ref{sec_field}
below and \cite{AcLuVo}).

The main result of this theory is that the time rescaling
induces a rescaling
\begin{equation}
a_{k} \longrightarrow \frac{1}{\lambda}e^{-i\frac{t}{\lambda^2}(\omega(k)-
\Omega)}a_{k}
\end{equation}
of the quantum field, defining the Hamiltonian (1), which in the present
paper will be assumed to be a scalar boson field: ($[a_k,a_{k'}]=\delta(k-k')$)
(the meaning of $\omega(k)$ and $\Omega$ will be described in next chapter)
and, in the limit $\lambda \rightarrow 0$, the rescaled field 
becomes a quantum white noise (or master field) $b_{\Omega}(t,k)$ 
satisfying the commutation relations
\begin{equation}
[b_{\Omega}(t,k),b^\dagger_{\Omega'}(t',k')]=\delta_{\Omega,\Omega'}
2\pi \delta(t-t')\delta(k-k')\delta(\omega(k)-\Omega).\label{9}
\end{equation}
Moreover, if the initial state of the field is a mean zero gauge
invariant Gaussian state $\rho_f(0)$ with correlations:
\begin{equation}
\langle a^\dagger_{k} a_{k'}\rangle = {N}(k)\delta(k-k')\label{10}
\end{equation}
then the state of the limit white noise will be of the same
type with correlations
\begin{subequations}
\label{11a}
\begin{equation}
\langle b^\dagger_{\Omega}(t,k) b_{\Omega'}(t',k')\rangle
=\delta_{\Omega,\Omega'}
2\pi \delta(t-t')\delta(k-k')\delta(\omega(k)-\Omega)N(k)
\end{equation} 
\begin{equation}
\langle b_{\Omega}(t,k) b^\dagger_{\Omega'}(t',k')\rangle
=\delta_{\Omega,\Omega'}
2\pi \delta(t-t')\delta(k-k')\delta(\omega(k)-\Omega)(N(k)+1).
\end{equation} 
\end{subequations}

It is now well understood that this scheme plays an
important role in the analysis of the limit (\ref{6}) when $X$ is a
system operator. In Sec.\ref{sec_field}, we describe a new development of the stochastic limit which allows to extend this scheme to a class of observables describing the slow degrees of freedom of the field.

The remaining part of this paper is arranged as follows:
In Sec. \ref{Sec.2} we consider a model which drives the
system to a non-equilibrium stationary state. It describes a quantum system
put between two reservoirs at different temperatures. By analysis of
the reduced density matrix with stochastic limit, we show that this
system has non-equilibrium 
stationary state which doesn't satisfy the detailed balance condition.
In Sec.\ref{sec_field}, we apply the stochastic limit
to the slow degrees of freedom of the field. 
This allows to define the currents associated to these degrees of freedom
and to discuss their properties. In terms of these currents,
we define the dynamical detailed balance condition
which is a generalization of the usual detailed balance condition.
In addition, we show that in the linear approximation these currents
satisfy the Onsager reciprocal relations \cite{Onsager31}.
In Sec.\ref{time symmetricity}, we investigate a master equation
characterization of this dynamical detailed balance condition,
which corresponds to the well-known fact that the usual detailed balance
condition is characterized by the master equation which drives the state
to equilibrium
\cite{Gorini-Kossokowski76,Kossakowski-Frigerio-Gorini-Verri77}.
Then in the next section, we introduce the local KMS
condition and prove that
it is equivalent to the dynamical detailed balance condition
for the state. In addition, we consider another model in which the system
interacts with an environment whose state is non-equilibrium
and satisfies the local KMS condition.
We show that such states of the environment drive
the system to a non-equilibrium state satisfying the local KMS condition
with a non linear temperature function which is uniquely determined
by the state of the field.
Finally in Sec.\ref{sec:sum}, we
summarize the contents of this paper and  discuss related topics.

\section{Deduction of the Stochastic Schr\"{o}dinger, Langevin and
master equation}\label{Sec.2}
In this section, we consider a model in which the system is driven to 
a non-equilibrium stationary state by its interaction with
two non-equilibrium boson fields. This interaction
is described by the  Hamiltonian
\begin{subequations}\label{eq:Hamiltonian1}
\begin{equation}
H=H_0+\lambda\sum_{j=1,2}H_{I_j},
\quad (\lambda ~\mbox{is a coupling constant.})
\end{equation}
\begin{equation}
H_0=H_S+H_B,\quad
H_S=\sum_{l}\epsilon_l |\epsilon_l\rangle\langle\epsilon_l|,\quad
H_B=\sum_{j}\int \omega_{j}(k)a^\dagger_{j,k}a_{j,k}
\quad [a_{j,k},a_{j',k'}^\dagger]=\delta_{jj'}\delta(k-k'), 
\end{equation}
\begin{equation}
H_{I_j}
=\int dk~\left(g_{j}(k)D_ja^\dagger_{j,k}+g^*_{j}(k)D_j^\dagger a_{j,k}\right)
\end{equation}
where $D_j$ and $D^\dagger_j$ are operators on the system space, $a_{j,k}$ and 
$a^\dagger_{j,k}$ are the annihilation and creation operators of the 
$j$-th field (j=1,2) and $g_{j}(k)$ is a form factor.
\end{subequations}

The initial state of each field is a Gibbs state at temperature $\beta_j^{-1}$
and chemical potential $\mu_j$ with respect to the free Hamiltonian
(throughout the present paper we assume $\omega_j(k)-\mu_j>0$ for all $k$ 
as usual), i.e. the mean zero gauge invariant Gaussian state with correlations:
\begin{equation}\label{correlations13}
\langle a_{j,k}^\dagger a_{j',k'} \rangle=
\delta_{jj'}N(k;\beta_j,\mu_j)\delta(k-k'),\quad
N(k;\beta_j,\mu_j)=\frac{1}{e^{\beta_j(\omega_j(k)-\mu_j)}-1}
\end{equation}
The Schr\"{o}dinger equation in the interaction picture is
\begin{equation}
\frac{d}{dt}U^{(\lambda)}_t=-i \lambda H_I(t) U^{(\lambda)}_t,\quad
U^{(\lambda)}_t=e^{i t H_0}e^{-i t H}
\end{equation}
where
\begin{subequations}
\begin{eqnarray}\label{eq:interaction_Hamiltonian}
H_I(t)&=&\sum_{j=1,2}e^{it H_0}H_{I_j}e^{-it H_0}
\nonumber\\
&=&\sum_{j=1,2}\sum_{\omega \in F}\sum_{lm}
\int dk~ \left(g_{j:lm} (k) E_\omega(lm) a_{j,k}^\dagger e^{+i(\omega_i(k)-\omega)t}+g^*_{j:lm} (k) E^\dagger_\omega(lm) a_{j,k}e^{-i(\omega_i(k)-\omega)t}\right)
\end{eqnarray}
\begin{equation}
g_{j:lm}(k)=g_j(k) \langle \epsilon_l|D|\epsilon_m \rangle,\quad
E_\omega(lm)=\sum_{\epsilon_r \in F_\omega}
\langle \epsilon_r-\omega |\epsilon_l\rangle\langle \epsilon_m|\epsilon_r\rangle |\epsilon_r-\omega \rangle\langle \epsilon_r|,\quad
\end{equation}
\begin{equation}
F= \{\omega=\epsilon_r-\epsilon_{r'}; \epsilon_r,\epsilon_{r'} \in 
{\rm Spec.} (H_S) \},\quad
F_{\omega}=\{\epsilon_{r'} \in {\rm Spec.} (H_S) ; \epsilon_{r'}-\omega \in {\rm Spec.} H_S\}
\end{equation}
\end{subequations}
In the following, for simplicity, we assume that $H_S$ is generic, i.e.\\
1) the spectrum Space $H_S$ is not degenerate\\
2) For any $\omega$ $|F_{\omega}|=1$, i.e. there exist a unique pair of energy
levels $\epsilon_l, \epsilon_m \in {\rm Spec.} (H_S)$ such that $\omega=\epsilon_m-\epsilon_l$\\
In such case, (\ref{eq:interaction_Hamiltonian}) becomes
\begin{subequations}
\begin{equation}
H_I(t)=\sum_{j=1,2}\sum_{\omega \in F}
\int dk~ \left(g_{j:\omega} (k) E_\omega a_{j,k}^\dagger e^{+i(\omega_i(k)-\omega)t}+g^*_{j:\omega} (k) E^\dagger_\omega a_{j,k}e^{-i(\omega_i(k)-\omega)t}\right)
\end{equation}
where
\begin{equation}
g_{j:\omega}(k)=g_j(k) \langle \epsilon_l|D_j|\epsilon_m \rangle,\quad
E_\omega=|\epsilon_l\rangle\langle \epsilon_m|,\quad \mbox{for}~\epsilon_l, \epsilon_m ~\mbox{s.t.}~ \epsilon_m-\epsilon_l=\omega.
\end{equation}
\end{subequations}

Giving an Hamiltonian such as (\ref{eq:Hamiltonian1}) the stochastic limit
technique proceeds in four steps:\\
\begin{enumerate}
\item Write the associated white noise Hamiltonian (WNH) equation
(\ref{eq:white_noise_eq1}).
\item The causally normally ordered form of the WNH equation gives
the Stochastic Schr\"{o}dinger (SS) equation (\ref{eq:SS}).
\item From the SS one deduces the Langevin equation 
(e.g. (\ref{eq:lange}) and (\ref{nama_def})).
\item Partial trace of the Langevin gives the master equation 
(e.g. (\ref{eq:master_equation})).
\end{enumerate}
In the following, we shall describe the results of these steps for our models
and we refer to \cite{AcLuVo} for a detailed description of the steps
necessary to achieve these results.

Applying stochastic limit as explained in Sec. \ref{sec:intro},
we obtain the white noise
Hamiltonian equation
\begin{subequations}\label{eq:white_noise_eq1}
\begin{equation}
\frac{d}{dt}U_t=-i \sum_{j=1,2}\sum_{\omega\in F}
\left(E_\omega b^\dagger_{t:j,\omega}+E^\dagger_\omega 
b_{t:j,\omega}
\right)U_t
\end{equation}
where
\begin{equation}
b_{t:j,\omega}=
\int dk~g^*_{j:\omega}(k)b_{t:j,\omega}(k),\quad
b_{t:j,\omega}(k)=\lim_{\lambda\rightarrow 0}\frac{1}{\lambda}
e^{-i(\omega_j(k)-\omega)t/\lambda^2}a_{j,k}.
\end{equation}
\end{subequations}
Notice that the state of the limit white noise will be of the same type
as (\ref{correlations13}) but with correlations
\begin{subequations}
\begin{eqnarray}
\langle b^\dagger_{t:j,\omega}(k) b_{t':j',\omega'}(k')\rangle
&=&\delta_{jj'}\delta_{\omega\omega'}2\pi\delta(t-t')\delta(k-k')\delta(\omega_j(k)-\omega)N(k;\beta_j,\mu_j)\\
\langle b_{t:j,\omega}(k) b^\dagger_{t':j',\omega'}(k')\rangle
&=&\delta_{jj'}\delta_{\omega\omega'}2\pi\delta(t-t')\delta(k-k')\delta(\omega_j(k)-\omega)\left(N(k;\beta_j,\mu_j)+1\right).
\end{eqnarray}
\end{subequations}
The SS equation associated to the WNH equation (\ref{eq:white_noise_eq1})
is
\begin{subequations}\label{eq:SS}
\begin{equation}
dU_t=-i\sum_{j=1,2}\left(E_{\omega}dB^\dagger_{t:j,\omega}+E^\dagger_\omega
dB_{t:j,\omega}
-i\left(\gamma_{-,j,\omega}E_\omega E_\omega^\dagger
+\gamma_{+,j,\omega}^*E^\dagger_\omega E_\omega\right)dt
\right)U_t.
\end{equation}
where
\begin{equation}
dB_{t:j,\omega}=\int^{t+dt}_{t}b_{\tau:j,\omega}d\tau,\quad
dB^\dagger_{t:j,\omega}=\int^{t+dt}_{t}b^\dagger_{\tau:j,\omega}d\tau
\end{equation}
are stochastic differentials and satisfy the Ito table
\begin{equation}
dB_{t:j,\omega} dB^\dagger_{t:j',\omega'}=2\delta_{jj'}
\delta_{\omega\omega'}{\rm Re}\gamma_{-,j,\omega}dt,\quad
dB^\dagger_{t:j,\omega} dB_{t:j',\omega'}=2\delta_{jj'}
\delta_{\omega\omega'}{\rm Re}\gamma_{+,j,\omega}dt
\end{equation}
\begin{equation}
dt dB_{t:j,\omega}=dB_{t:j,\omega}dB_{t:j',\omega'}
=dB^\dagger_{t:j,\omega}dB^\dagger_{t:j',\omega'}=dt dB^\dagger_{t:j,\omega}=0.
\end{equation}
\end{subequations}
The main physical information is contained in the generalized susceptivities
(or transport coefficients):
\begin{subequations}
\begin{eqnarray}
\gamma_{-,j,\omega}&=&\int dk |g_{j,\omega}(k)|^2\frac{-i(N(k;\beta_j,\mu_j)+1)}{\omega-\omega_j(k)-i0}\nonumber\\
&=&\pi \int dk~|g_{j,\omega}(k)|^2\frac{e^{\beta_i(\omega_j(k)-\mu_j)}}{e^{\beta_j(\omega_j(k)-\mu_j)}-1}\delta(\omega_j(k)-\omega)-i{\rm P.P}\int dk~
\frac{|g_{j,\omega}(k)|^2}{\omega_j(k)-\omega}\frac{e^{\beta_i(\omega_j(k)-\mu_j)}}{e^{\beta_j(\omega_j(k)-\mu_j)}-1}
\\
\gamma_{+,j,\omega}&=&\int dk |g_{j,\omega}(k)|^2\frac{-iN(k;\beta_j,\mu_j)}{\omega-\omega_j(k)-i0}\nonumber\\
&=&\pi \int dk~|g_{j,\omega}(k)|^2\frac{1}{e^{\beta_j(\omega_j(k)-\mu_j)}-1}\delta(\omega_j(k)-\omega)-i{\rm P.P}\int dk~
\frac{|g_{j,\omega}(k)|^2}{\omega_j(k)-\omega}\frac{1}{e^{\beta_j(\omega_j(k)-\mu_j)}-1}.
\end{eqnarray}
\end{subequations}
For an operator $X$ of the system space ${\cal H}_S$,
from the SS equation (\ref{eq:SS}),
one obtains the Langevin equation
\begin{eqnarray}\label{eq:lange}
d\left(U_t^\dagger X U_t\right)=&&
\sum_{j=1,2}\sum_{\omega\in F}\left[i
\left(U_t^\dagger [E_\omega,X] U_t dB^\dagger_{t,j,\omega}
+U_t^\dagger [E^\dagger_\omega,X] U_t dB_{t,j,\omega}\right.\right.
\nonumber\\
&&
\left.-{\rm Im}
\gamma_{-,j,\omega}U_t^\dagger [E_\omega E_\omega^\dagger,X]U_t~dt
+{\rm Im}
\gamma_{+,j,\omega}U_t^\dagger [E^\dagger_\omega E_\omega,X]U_t~dt
\right)
\nonumber\\
&&\left.-U_t^\dagger\left({\rm Re} \gamma_{-,j,\omega}
\left(\{E_\omega E_\omega^\dagger,X\}-2E_\omega X E^\dagger_\omega\right)
+{\rm Re} \gamma_{+,j,\omega}
\left(\{E^\dagger_\omega E_\omega,X\}-2E^\dagger_\omega X E_\omega\right)
\right)U_t~dt\right].
\end{eqnarray}
The Langevin equation with the state for some operators of the field degrees
of freedom will be discussed in the following section (see (\ref{nama_def})).
Taking the partial expectation value of both sides of
this Langevin equation with the state (\ref{correlations13}),
the master equation for the reduced density matrix is obtained:
\begin{subequations}\label{eq:master_equation}
\begin{eqnarray}\label{eq:master_equation_a}
\frac{d}{dt}\rho_S(t)&=&-i[\Delta,\rho_S(t)]\nonumber\\
&&-\sum_{\omega\in F}\Gamma_{-,\omega}\left(\frac{1}{2}\left\{E^\dagger_\omega E_\omega,\rho_S(t)\right\}
-E_\omega\rho_S(t)E_\omega^\dagger\right)
-\sum_{\omega\in F}
\Gamma_{+,\omega}\left(\frac{1}{2}
\left\{E_\omega E^\dagger_\omega,\rho_S(t)\right\}
-E_\omega^\dagger\rho_S(t)E_\omega\right)
\end{eqnarray}
\begin{eqnarray}
&&\Delta=i~\sum_{\omega\in F}\sum_{j=1,2}\left({\rm Im}(\gamma_{-,j\omega})
E^\dagger_\omega E_\omega-{\rm Im}(\gamma_{+,j\omega})E_\omega E_\omega^\dagger\right)\\
&&\Gamma_{\mp,\omega}=2{\rm Re}~
\sum_{j=1,2}\gamma_{\mp,j,\omega} \ge 0,\quad (\Gamma_{\mp,\omega}=0 \quad \mbox{for} \quad \omega \le 0).
\end{eqnarray}
\end{subequations}
The generator of (\ref{eq:master_equation_a})
has the standard GKSL form \cite{GKSL}.
For the off-diagonal matrix elements $\rho_{mn}(t)=\langle\epsilon_m|\rho_S(t)|\epsilon_n\rangle$ $(m\neq n)$ we obtain 
\begin{subequations}\label{eq:offdiagonal_equation}
\begin{eqnarray}
\frac{d}{dt}\rho_{mn}(t)&=&\left(i\Delta_{mn}-G_{mn}\right)\rho_{mn}(t)
\end{eqnarray}
\begin{equation}
\Delta_{mn}=
\sum_{l}\left(\theta_{-,\epsilon_m-\epsilon_l}
-\theta_{-,\epsilon_n-\epsilon_l}
-\theta_{+,\epsilon_l-\epsilon_m}
+\theta_{+,\epsilon_l-\epsilon_n}
\right),\quad \theta_{\mp,\omega}={\rm Im} \gamma_{\mp,j,\omega}
\end{equation}
\begin{equation}\label{eq:def_G_MN}
G_{mn}=\sum_{l}\left(\Gamma_{ml}+\Gamma_{nl}\right)>0,\quad
\mbox{where}\quad \Gamma_{ml}=\left\{
\begin{array}{ccc}
\Gamma_{-,\epsilon_m-\epsilon_l}&\mbox{for}&\epsilon_m > \epsilon_l\\
\Gamma_{+,\epsilon_l-\epsilon_m}&\mbox{for}&\epsilon_m < \epsilon_l
\end{array}\right.
\end{equation}
\end{subequations}
which shows that these elements vanish at $t\rightarrow \infty$
whenever $G_{mn}\neq 0,~(\forall ~m,n)$.

The diagonal matrix elements $\rho_{mm}(t)$ describe a classical birth and
death process characterized by the equation
\begin{subequations}\label{eq:diagonal_equation}
\begin{eqnarray}
\frac{d}{dt}\rho_{mm}(t)&=&-\sum_{l}\left(
(\Gamma_{-,\epsilon_m-\epsilon_l}+\Gamma_{+,\epsilon_l-\epsilon_m})
\rho_{mm}(t)-
(\Gamma_{-,\epsilon_l-\epsilon_m}+\Gamma_{+,\epsilon_m-\epsilon_l})
\rho_{ll}(t)\right)\nonumber \\
&=& -\sum_l\left(\Gamma_{ml}~\rho_{mm}(t)-
\Gamma_{lm}~\rho_{ll}(t)\right)\\
&=& -\sum_{l}A_{ml}\rho_{ll},\quad A_{ml}=\left\{
\begin{array}{cc}
\sum_{l}\Gamma_{ml} & \mbox{for}\quad l=m\\
-\Gamma_{lm}& \mbox{for}\quad l\neq m
\end{array}\right.
\end{eqnarray}
\begin{equation}
\frac{\Gamma_{ml}}{\Gamma_{lm}}=
\frac{{\rm Re}\sum_{j=1,2}\gamma_{-,j,\epsilon_m-\epsilon_l}}
{{\rm Re}\sum_{j=1,2}\gamma_{+,j,\epsilon_m-\epsilon_l}} \quad (\mbox{for} \quad \epsilon_m>\epsilon_l), \quad \mbox{or}\quad 
\frac{\Gamma_{ml}}{\Gamma_{lm}}=
\frac{{\rm Re}\sum_{j=1,2}\gamma_{+,j,\epsilon_l-\epsilon_m}}{{\rm Re}\sum_{j=1,2}\gamma_{-,j,\epsilon_l-\epsilon_m}} \quad (\mbox{for} \quad \epsilon_l>\epsilon_m).
\end{equation}
\end{subequations}
Notice that this quotient is universal in the sense that it does not depend on 
$g_j$ whenever in the interaction (\ref{eq:Hamiltonian1}) form factors $g_i$
do not depend on $j$ ($g_j=g$). 
When the matrix $A$ has a non-trivial eigenvector associated to 
the $0$ eigenvalue, a stationary state exists. In addition,
the convergence to the stationary state from any initial state
$\rho_S(0)$ is guaranteed under quite general conditions 
(cf. \cite{Accardi-Fagnola}). 
Notice, that 
the stationary solution of (\ref{eq:master_equation}) satisfies the detailed
balanced condition, i.e.
\begin{equation}\label{db_condition0}
\frac{\rho_{mm}}{\rho_{ll}}=\frac{\Gamma_{lm}}{\Gamma_{ml}},
\end{equation}
if and only if the coefficients $\Gamma_{ml}$ satisfy
\begin{equation}\label{db_condition}
\frac{\Gamma_{ml}}{\Gamma_{lm}}=\frac{\Gamma_{mk}}{\Gamma_{km}}\frac{\Gamma_{kl}}{\Gamma_{lk}},\quad \forall ~m,l,k
\end{equation}
In the non-equilibrium case (\ref{db_condition}) is not satisfied.
With this model given by (\ref{eq:Hamiltonian1}) and (\ref{correlations13}),
(\ref{db_condition}) is satisfied only in some special cases
(for example when both fields have the same temperature and
chemical potential, or the system has only one Bohr frequency).

In general, the stationary state of the master equation (\ref{eq:master_equation}) can be described by the nonlinear temperature function
\begin{equation}\label{eq:gen_temp1}
\beta_S(\epsilon_m)=\frac{-1}{\epsilon_m}\log\rho_{mm} >0
\end{equation}
as
\begin{equation}\label{eq:gen_temp2}
\rho_S=\frac{e^{-\beta_S(H_S)H_S}}{Z},\quad 
Z={\rm tr}_S\left({e^{-\beta_S(H_S)H_S}}\right).
\end{equation}
The state is Gibbs state the function $\beta_S(H_S)$ becomes constant.
This fact actually leads to the idea that a rather wide class of
non-equilibrium
stationary states can be treated with such generalized temperature functions.
This notion is valid not only for the system but also for the state of the 
field. Indeed, in the Sec.\ref{ModelB} below, we will consider another
model in which the system is driven to a non-equilibrium stationary state by an
interaction with a non-equilibrium field described by a generalized temperature function.

\section{Microscopic currents and dynamical Detailed Balance}\label{sec_field}
In the previous section, we have investigated the dynamics of a
system interacting
with fields in a non-equilibrium situation and we have already remarked some
important difference from the equilibrium case. However one can see a
more direct and crucial difference through the study of the dynamics of the 
field degrees of freedom.

\subsection{Slow degrees of freedom and micro-current}
In order to investigate the dynamics of the field, it is important to notice
that some operators of the field degrees of freedom, i.e. the slow degrees,
survive even after stochastic limit. As we explained
in the introduction, the rescaled field operators
$a_k$ and $a_k^\dagger$ become
white noise operators denoted by 
$b_{\omega}(t,k)$ and $b_{\omega}^\dagger(t,k)$ whose
commutation relation is given by (\ref{9}). 
Due to this fact we can intuitively say that the fast degrees of the field 
become noise (singular) in the stochastic limit. However
we can describe the time evolution of some of operators of the field
in terms of the rescaled time even after stochastic limit, and
this approach gives us meaningful information on the original
dynamics as well as on the system operator. Since the stochastic limit is an
asymptotic theory, mathematically we have to prove the convergence of the 
dynamics and this has been done elsewhere \cite{AcLuVo}.
In the present paper, we apply the theory to the number operator
in the model and discuss its physical meaning.

Let us sketch how to compute the time evolution of the number operator 
$n_k=a^\dagger_k a_k$ under the white noise equation
\begin{equation}
\frac{d}{dt}U_t=-i\sum_{\omega\in F}\left(E_\omega b^\dagger_t+E^\dagger_\omega b_t\right)U_t.
\end{equation}
We will illustrate the calculation only in the simplest (Fock) case.
The more general states (\ref{10}) can be reduced to a linear combination
of two independent Fock representations (cf. \cite{AcLuVo} section 2.18).
The key formula to apply stochastic limit to the number operator of
the field $n_k=a^\dagger_k a_k$ is
\begin{equation}
[b_{\omega}(t,k), n_{k'}]=b_{\omega}(t,k)\delta(k-k').
\end{equation}
The Heisenberg evolution of $n_k$, after the stochastic limit is described
by the Langevin equation
\begin{eqnarray}\label{nama_def}
\frac{d}{dt}\left(U^\dagger_t n_k U_t\right)
&=&i\sum_{\omega \in F}
U^\dagger_t [E_\omega b^\dagger_{t,\omega}
+E^\dagger_\omega b_{t,\omega},~n_k]U_t \nonumber\\
&=&-i \sum_{\omega \in F}\left(U^\dagger_t
\left(E_\omega b^\dagger_{\omega}(t,k)-E_\omega^\dagger b_{\omega}(t,k)\right)
U_t\right)\nonumber\\
&=&-i \sum_{\omega \in F}
\left(b^\dagger_{\omega}(t,k)U_t^\dagger E_\omega U_t
-U_t^\dagger E_\omega U_t b_{\omega}(t,k)
+[U_t^\dagger,b^\dagger_{\omega}(t,k)]E_\omega U_t
-U^\dagger_tE^\dagger_\omega [b_{\omega}(t,k),U_t]\right)
\nonumber\\
&=&-i \sum_{\omega \in F}
\left(b^\dagger_{\omega}(t,k)U_t^\dagger E_\omega U_t
-U_t^\dagger E_\omega U_t b_{\omega}(t,k)\right. \nonumber\\
&& \left.
+i\left(
\gamma_\omega(k)U^\dagger_t E^\dagger_\omega E_\omega U_t
+\gamma_\omega(k)U^\dagger_t E^\dagger_\omega E_\omega U_t
\right)\right)\delta(\omega(k)-\omega)
\end{eqnarray}
where in the Fock case 
\begin{equation}\label{Fock}
\gamma_\omega(k)= \pi |g_\omega(k)|^2. 
\end{equation}
Taking partial trace $\langle~\cdot~\rangle$
over the initial state of the system and noise we obtain the
evolution equation of the mean number of quanta
\begin{equation}
\frac{d}{dt}\langle U^\dagger_t n_k U_t \rangle
=2\sum_{\omega\in F}\gamma_\omega(k)\langle U^\dagger_t E^\dagger_\omega E_\omega U_t \rangle\delta(\omega(k)-\omega).
\end{equation}
This can be expressed in terms of the time evolution
(under the master equation (\ref{eq:master_equation})) of the reduced density
matrix $\rho_S(t)$, i.e.
\begin{eqnarray}\label{eq:numberB0}
\frac{d}{dt}\langle U^\dagger_t n_k U_t \rangle
&=&2\sum_{\omega\in F}
\delta(\omega(k)-\omega)\gamma_\omega(k)
{\rm tr}_S\Big(E^\dagger_{\omega} E_\omega \rho_S(t)\Big)
\nonumber\\
&=&2\sum_{\epsilon_m > \epsilon_n}
\delta(\omega(k)-(\epsilon_{m}-\epsilon_n)){\rm tr}_S\Big(
\gamma_\omega(k)|\epsilon_m\rangle\langle\epsilon_m|\rho_S(t)\Big)
\end{eqnarray}
In the case of a general initial state described by
(\ref{10}), the computation is similar, and one get
\begin{subequations}\label{eq:numberB1}
\begin{eqnarray}
\frac{d}{dt}\langle U^\dagger_t n_k U_t \rangle
&=&2\sum_{\omega\in F}
\delta(\omega(k)-\omega){\rm tr}_S\left(
\left(\gamma_{-,\omega}(k) E^\dagger_{\omega} E_\omega 
-\gamma_{+,\omega}(k)E_\omega E_\omega^\dagger\right)\rho_S(t)\right)
\nonumber\\
&=&2\sum_{\epsilon_m > \epsilon_n}
\delta(\omega(k)-(\epsilon_{m}-\epsilon_n))
{\rm tr}_S\left(
\Big(\gamma_{-,\epsilon_{m}-\epsilon_n}(k)|\epsilon_m\rangle\langle\epsilon_m|
-\gamma_{+,\epsilon_{m}-\epsilon_n}(k)|\epsilon_n\rangle\langle\epsilon_n|
\Big)\rho_S(t)\right)
\end{eqnarray}
where now instead of $\gamma_{\omega}(k)$ given by (\ref{Fock}) one has:
\begin{equation}
\gamma_{-,\omega}(k)=\gamma_\omega(k)(N(k)+1),\quad 
\gamma_{+.\omega}(k)=\gamma_\omega(k)N(k).
\end{equation}
\end{subequations}
As the consequence of (\ref{eq:numberB0}), once we obtain the time evolution of $\rho_S(t)$ by solving the master equation discussed in the previous section,
we find the time evolution of the number operator of field degrees of freedom.

In order to apply (\ref{eq:numberB0}) to the model discussed in the previous section, we can consider the number operator for each field.
Defining $n_{j,k}=a^\dagger_{j,k}a_{j,k}$, we obtain
\begin{subequations}
\label{eq:numberA1}
\begin{eqnarray}
\frac{d}{dt}\langle U^\dagger_t n_{j,k} U_t \rangle
&=&2\sum_{\omega\in F}
\delta(\omega_j(k)-\omega){\rm tr}_S\left(
\left(\gamma_{-,\omega,j}(k) E^\dagger_{\omega} E_\omega 
-\gamma_{+,\omega,j}(k)E_\omega E_\omega^\dagger\right)\rho_S(t)\right)
\nonumber\\
&=&2\sum_{\epsilon_m > \epsilon_n}
\delta(\omega_j(k)-(\epsilon_m-\epsilon_n))
{\rm tr}_S\left(
\Big(\gamma_{-,\epsilon_m-\epsilon_n,j}(k)|\epsilon_m\rangle\langle\epsilon_m|
-\gamma_{+,\epsilon_m-\epsilon_n,j}|\epsilon_n\rangle\langle\epsilon_n|\Big)
\rho_S(t)\right)\nonumber\\
&=&2\sum_{\epsilon_m > \epsilon_n}
\delta(\omega_j(k)-(\epsilon_m-\epsilon_n))
\Big(\gamma_{-,\epsilon_m-\epsilon_n,j}(k)\rho_{mm}(t)
-\gamma_{+,\epsilon_m-\epsilon_n,j}\rho_{nn}(t)\Big)
\end{eqnarray}
where
\begin{equation}
\gamma_{-,\omega,j}(k)=\pi|g_{j,\omega}(k)|^2~(N(k;\beta_j,\mu_j)+1),\quad
\gamma_{+,\omega,j}(k)=\pi|g_{j,\omega}(k)|^2~N(k;\beta_j,\mu_j).
\end{equation}
\end{subequations}
This time dependence of the slow of degrees of freedom of the
field is due to the interaction with the system and is a direct evidence of the existence of a family of
currents passing through the system: one for each proper frequency $\omega=\varepsilon_m-
\varepsilon_n>0$. To investigate these currents, let us define, for each $\epsilon_m >\epsilon_n$,
the region $\Omega_{mn}$ in $k$-space, resonating with the frequency $\omega_{mn}:=\varepsilon_m-
\varepsilon_n$ which includes all $k_{mn}$ such that
\begin{equation}
\omega(k_{mn})-(\epsilon_m-\epsilon_n)=0.\label{37d}
\end{equation}
Then define the microscopic number current, associated to the frequency $\omega_{mn}$ by:
\begin{subequations}\label{currentA}
\begin{eqnarray}
J_{j,mn}(t):&=&\frac{d}{dt}\Big(\int_{\Omega_{mn}} 
dk ~\langle U_t^\dagger n_{j,k} U_t \rangle  \Big)\nonumber \\
&=&2\left({\rm Re}\gamma_{-,j,\epsilon_m-\epsilon_n}~\rho_{mm}(t)
-{\rm Re}\gamma_{+,j,\epsilon_m-\epsilon_n}~\rho_{nn}(t)\right),\nonumber
\\
J_{j,mn}:&=&2\left({\rm Re}\gamma_{-,j,\epsilon_m-\epsilon_n}~\rho_{mm}
-{\rm Re}\gamma_{+,j,\epsilon_m-\epsilon_n}~\rho_{nn}\right),\quad
\mbox{(in stationary state of the system)}
\nonumber\\
&=& 2 \gamma_{j,mn}~\rho_{mm}\frac{e^{(\epsilon_m-\epsilon_n-\mu_j)\beta_j}}
{e^{(\epsilon_m-\epsilon_n-\mu_j)\beta_j}-1}\left(1-
e^{-(\epsilon_m-\epsilon_n-\mu_j)\beta_j}\frac{\rho_{nn}}{\rho_{mm}}\right)
\label{currentANS}
\end{eqnarray}
and similarly the microscopic energy current
\begin{eqnarray}
J^{E}_{j,mn}(t):&=&\frac{d}{dt}\Big(\int_{\Omega_{mn}} 
dk ~\omega_j(k)\langle U_t^\dagger n_{j,k} U_t \rangle  \Big)\nonumber \\
&=&2\left(\epsilon_m-\epsilon_n\right)\left({\rm Re}\gamma_{-,j,\epsilon_m-\epsilon_n}~\rho_{mm}(t)
-{\rm Re}\gamma_{+,j,\epsilon_m-\epsilon_n}~\rho_{nn}(t)\right)\nonumber
\\
J^E_{j,mn}:&=&2\left(\epsilon_m-\epsilon_n\right)
\left({\rm Re}\gamma_{-,j,\epsilon_m-\epsilon_n}~\rho_{mm}
-{\rm Re}\gamma_{+,j,\epsilon_m-\epsilon_n}~\rho_{nn}\right),\quad
\mbox{(in stationary state of the system)}\nonumber\\
&=& 2 (\epsilon_m-\epsilon_n) \gamma_{j,mn}~\rho_{mm}
\frac{e^{(\epsilon_m-\epsilon_n-\mu_j)\beta_j}}
{e^{(\epsilon_m-\epsilon_n-\mu_j)\beta_j}-1}\left(1-
e^{-(\epsilon_m-\epsilon_n-\mu_j)\beta_j}\frac{\rho_{nn}}{\rho_{mm}}\right)
\label{currentAES}
\end{eqnarray}
where 
\begin{equation}
\gamma_{j,mn}=\pi \int_{k\in \Omega_{mn}} dk ~|g_{j,\epsilon_m-\epsilon_n}(k)|\delta(\omega_j(k)-(\epsilon_m-\epsilon_n)).
\end{equation}
\end{subequations}
The term {\it microscopic\/} here refers to the fact that we define one current for each atomic frequency. 
We see, from (\ref{currentA}) that in the stationary state for the system
$$\rho_S(t)=\rho_S$$
we have a constant flow of quanta from the field to the system.

The sum, over all $m$ and $n$, of our micro-currents gives two macro-currents
which coincide with those defined by H. Spohn and J. L. Lebowitz
in terms of the master equation \cite{Spohn-Lebowits78}.
In fact, as seen in (\ref{currentA}), these currents can be represented with
the matrix elements of the reduced density matrix and the generators
of master equation like they defined (cf. also the formulas (\ref{eq41})
and (\ref{eq42}) bellow).
However
the micro-currents are essential to define dynamical detailed balance
and the fact that
we started from the dynamics of the fields and deduced them
gives a physical interpretation to these currents.

Moreover our approach shows that in fact
a much stronger condition is satisfied namely:
for each Bohr frequency $\omega\in F$ the mean micro-current relative to
the frequency $\omega=\varepsilon_m-\varepsilon_n$ is constant.
This means that, for each $\omega\in F$, the flow of quanta from the modes of the field resonating with the frequency $\omega$
(in the sense of condition (\ref{37d})) is constant. 
Thus the current of quanta in the field is split into a family of independent {\it microscopic currents\/}, one for each Bohr frequency $\omega$. In the stationary state each of these microscopic currents is constant: we shall call this fact {\it dynamical detailed balance\/}.
This condition gives a simple and experimentally measurable relation
expressing the microcurrent associated to a transition between
two levels $\epsilon_m\rightarrow\epsilon_n$ as a linear combination of
the occupation probabilities of the two levels, with coefficients given
by the generalized susceptivities (transport coefficients).

The usual (equilibrium) detailed balance condition
is the particular case of the dynamical one corresponding to the case
in which all the microscopic currents are zero. In fact in this case
equation (\ref{currentA}) is reduced to
$$
{\rho_{nn}\over\rho_{mm}}\,=\frac{{\rm tr}\Big(|n\rangle\langle n|
\rho_S\Big)}{{\rm tr}\Big(|m\rangle\langle m|\rho_S\Big)}
=e^{\beta_j(\epsilon_m-\epsilon_n-\mu_j)},\quad \forall j=1,2
$$
for any $k_{mn}$ satisfying condition (\ref{37d}).
From this, by standard arguments, it follows that there exists a constant 
$\beta>0$ such that
$$\beta_1=\beta_2=\beta,\quad
\rho_{mm}={e^{-\beta\epsilon_m}\over Z_\beta}\,;\ Z_\beta=\sum_m e^{-\beta \epsilon_m}$$
so that $\rho_S$ is the Gibbs distribution.

In the general case the dynamical detailed balance condition is
\begin{equation}
2\left({\rm Re}\gamma_{-,\epsilon_m-\epsilon_n,j}~\rho_{mm}
-{\rm Re}\gamma_{+,\epsilon_m-\epsilon_n,j}~\rho_{nn}\right)=J_{j,mn}
\label{eqrho}
\end{equation}
This gives, for $m>0$
$$2{\rm Re}\gamma_{-,\epsilon_m-\epsilon_0,j}~\rho_{mm}
=2{\rm Re}\gamma_{+,\epsilon_m-\epsilon_0,j}~\rho_{00}+J_{j,m0}$$
or
$$\rho_{mm}=\frac{{\rm Re}\gamma_{+,\epsilon_m-\epsilon_0,j}}
{{\rm Re}\gamma_{-,\epsilon_m-\epsilon_0,j}}~\rho_{00}
+\frac{J_{j,m0}}{2{\rm Re}\gamma_{-,\epsilon_m-\epsilon_0,j}}$$
Replacing this into (\ref{eqrho}) we find
\begin{eqnarray}
&&2{\rm Re} \gamma_{-,\epsilon_m-\epsilon_n,j}\left[{{\rm Re}
\gamma_{+,\epsilon_m-\epsilon_0,j}\over
{\rm Re}\gamma_{-,\epsilon_m-\epsilon_0,j}}\,\rho_{00}+
{J_{j,m0}\over 2{\rm Re}\gamma_{-,\epsilon_m-\epsilon_0,j}}\right]
\nonumber\\
&&\hspace*{3cm}
=2{\rm Re}\gamma_{+,\epsilon_m-\epsilon_n,j}
\left[{{\rm Re}\gamma_{+,\epsilon_n-\epsilon_0,j}\over
{\rm Re}\gamma_{-,\epsilon_n-\epsilon_0,j}}\,\rho_{00}+
{J_{j,n0}\over 2{\rm Re}\gamma_{-,\epsilon_n-\epsilon_0,j}}
\right]+J_{j,mn}
\nonumber
\end{eqnarray}
or equivalently
\begin{eqnarray}
J_{j,mn}=&&2\left[{{\rm Re}\gamma_{-,\epsilon_m-\epsilon_n,j}
{\rm Re}\gamma_{+,\epsilon_m-\epsilon_0,j}\over{\rm Re}\gamma_{-,\epsilon_m-
\epsilon_0,j}}\,-{{\rm Re}\gamma_{+,\epsilon_m-\epsilon_n,j}
{\rm Re}\gamma_{+,\epsilon_n-\epsilon_0,j}\over
{\rm Re}\gamma_{-,\epsilon_n-\epsilon_0,j}}\right]
\rho_{00}\nonumber\\
&&\hspace*{4cm}
+{{\rm Re}\gamma_{-,\epsilon_m-\epsilon_n,j}\over
{\rm Re}\gamma_{-,\epsilon_m-\epsilon_0,j}}\,J_{j,m0}
-{{\rm Re}\gamma_{+,\epsilon_m-\epsilon_n,j}\over
{\rm Re}\gamma_{-,\epsilon_n-\epsilon_0,j}}\,J_{j,n0}
\end{eqnarray}
which shows that, under the dynamical detailed balance condition, the intensities of the microscopic currents are uniquely determined by the single sequence 
$J_{j,m0}$.

The following identities make the physical 
meaning of the currents $J_{j,mn}(t)$ and $J^E_{j,mn}(t)$ clear:
\begin{eqnarray}\label{eq41}
J_m(t):&=&\sum_{j=1,2}\left(\sum_{n<m}J_{j,mn}(t)-\sum_{n>m}J_{j,nm}(t)\right)
\nonumber\\
&=&-\frac{d}{dt}{\rm tr}\Big(|\epsilon_m\rangle\langle\epsilon_m|\rho_{S}(t)
\Big)
\end{eqnarray}
is the difference between the quanta emitted from and absorbed by the level
$\epsilon_m$.
\begin{eqnarray}\label{eq42}
\sum_{m}J^{E}_m(t):&=&\sum_{m}\sum_{j=1,2}\left(\sum_{n<m}
J^E_{j,mn}(t)-\sum_{n>m}J^E_{j,nm}(t)\right)
\nonumber\\
&=&
-\frac{d}{dt}{\rm tr}\Big(H_S\rho_{S}(t)\Big)
\end{eqnarray}
expresses the fact that
the variation of energy of the system is exactly balanced.

On the other hand, the behavior of each microscopic current $J_{j,mn}$
doesn't always follow a naive intuition.
For example, 
even in the symmetric configuration of interaction ($g_1(k)=g_2(k)=g(k)$ and
$\mu_1=\mu_2$), there are cases when some micro currents flow backward
(i.e. from the low to the high temperature reservoir), however it is impossible
that all micro currents flow backward.
A sufficient condition that the total energy current
\begin{equation}
J_{1}^{(E)}=\sum_{m}\left(\sum_{n<m}J^E_{1,mn}-\sum_{n>m}J^E_{1,nm}
\right)=-J_{2}^{(E)}
\end{equation}
is positive when the reservoir $1$ is at lower temperature than $2$
is that
\begin{equation}\label{state_futuu}
\frac{\rho_{mm}}{\rho_{nn}} <1, \quad \forall m > n,
\end{equation}
i.e. that there is no inversely populated state.
In addition, if all $J_{1,mn}$ and $J_{2,mn}$ have opposite sign,
the following strong relation ({\it Gibbs domination bound}) holds:
\begin{equation}
e^{-(\epsilon_m-\epsilon_n-\mu_1)\beta_1}\le
\frac{\rho_{mm}}{\rho_{nn}}\le
e^{-(\epsilon_m-\epsilon_n-\mu_2)\beta_2},\quad \epsilon_m >\epsilon_n.
\end{equation}
However
\begin{equation}\label{dynamical_db_condition}
J^{(E)}_{1,mn}=-J^{(E)}_{2,mn}
\end{equation}
is not true when the stationary state of the system does not satisfy
the detailed balance condition (See (\ref{eq:diagonal_equation}), 
(\ref{db_condition0}) and (\ref{db_condition})). In fact
\begin{subequations}\label{notdynamical_db_condition}
\begin{eqnarray}
J_{1,mn}+J_{2,mn}&=&\sum_{j=1,2}J_{j,mn}\nonumber\\
&=&2\sum_{j=1,2}\left({\rm Re}\gamma_{-,j,\epsilon_m-\epsilon_n}~\rho_{mm}
-{\rm Re}\gamma_{+,j,\epsilon_m-\epsilon_n}~\rho_{nn}\right)\nonumber\\
&=&\Gamma_{mn}\rho_{mm}-\Gamma_{nm}\rho_{nn}\neq 0,\\
J^E_{1,mn}+J^E_{2,mn}&\neq& 0.
\end{eqnarray}
\end{subequations}
In other words, these stationary current can satisfy 
(\ref{dynamical_db_condition}) if and only if 
the stationary state of the system satisfies the detailed balance
condition. When the stationary state can be described with detailed balance
condition, the generalized temperature defined by (\ref{eq:gen_temp2})
becomes constant which can be interpreted as the local temperature of
the system in between two fields. Thus
this condition gives a characterization of those non-equilibrium
stationary states which are local equilibrium stationary states with current. 
We show an important example of such state in the following, however
apart from few trivial cases, to satisfy the detailed balance condition 
strictly is impossible in this model as explained in the previous section.
We consider the case where the detailed balance condition is
satisfied approximately, i.e. the linear transport regime. 

\subsection{Linear approximation, local equilibrium and Onsager relation}
\label{subSec.linear_app}

Here we show that the stationary current defined by 
(\ref{currentA}) is consistent with well-known non-equilibrium physics
in linear regime.  
First we assume that the form factors in the
interactions are the same for the two fields ($g_1(k)=g_2(k):=g(k)$).
This implies that the stationary solution is  symmetric 
with respect to the indices 1 and 2.
Now consider a small variation of these parameters
\begin{equation}
\beta_0=\frac{\beta_1+\beta_2}{2},\quad
\delta\beta=\beta_1-\beta_2,\quad \mbox{and}\quad
\mu_0=\frac{\mu_1+\mu_2}{2},\quad
\delta\mu=\mu_2-\mu_1
\end{equation}
and the first order expansion of the stationary solution in
$\delta\beta$ and $\delta\mu$. This gives
\begin{equation}\label{expansion_1st}
\begin{array}{cc}
{\rho_{mm}}_{\small\left|
\begin{array}{cc}
\beta_1=\beta_0+\frac{\delta\beta}{2}, & \mu_1=\mu_0-\frac{\delta\mu}{2}\\
\beta_2=\beta_0-\frac{\delta\beta}{2}, & \mu_2=\mu_0+\frac{\delta\mu}{2}
\end{array}\right.}=&
\left(1+\frac{\delta\beta}{2}\frac{\partial}{\partial\beta_1}
-\frac{\delta\beta}{2}\frac{\partial}{\partial\beta_2}
-\frac{\delta\mu}{2}\frac{\partial}{\partial\mu_1}
+\frac{\delta\mu}{2}\frac{\partial}{\partial\mu_2}\right)
{\rho_{mm}}_{\small\left|
\begin{array}{cc}
\beta_1=\beta_0, & \mu_1=\mu_0\\
\beta_2=\beta_0, & \mu_2=\mu_0
\end{array}\right.}\\
&\\
&+ \mbox{higher order corrections}
\end{array}
\end{equation}
Using the symmetry (in $1,2$) of $\rho_{mm}$ at $\delta\beta=\delta\mu=0$:
\begin{equation}
\frac{\partial\rho_{mm}}{\partial \beta_1}_{\small\left|
\begin{array}{cc}
\beta_1=\beta_0, & \mu_1=\mu_0\\
\beta_2=\beta_0, & \mu_2=\mu_0
\end{array}\right.}
=
\frac{\partial\rho_{mm}}{\partial \beta_2}_{\small\left|
\begin{array}{cc}
\beta_1=\beta_0, & \mu_1=\mu_0\\
\beta_2=\beta_0, & \mu_2=\mu_0
\end{array}\right.},\quad
\frac{\partial\rho_{mm}}{\partial \mu_1}_{\small\left|
\begin{array}{cc}
\beta_1=\beta_0, & \mu_1=\mu_0\\
\beta_2=\beta_0, & \mu_2=\mu_0
\end{array}\right.}
=
\frac{\partial\rho_{mm}}{\partial \mu_2}_{\small\left|
\begin{array}{cc}
\beta_1=\beta_0, & \mu_1=\mu_0\\
\beta_2=\beta_0, & \mu_2=\mu_0
\end{array}\right.}
\end{equation}
all the cross terms in (\ref{expansion_1st}) cancel and we obtain
\begin{equation}
{\rho_{mm}}_{\small\left|
\begin{array}{cc}
\beta_1=\beta_0+\frac{\delta\beta}{2}, & \mu_1=\mu_0-\frac{\delta\mu}{2}\\
\beta_2=\beta_0-\frac{\delta\beta}{2}, & \mu_2=\mu_0+\frac{\delta\mu}{2}
\end{array}\right.}=
{\rho_{mm}}_{\small\left|
\begin{array}{cc}
\beta_1=\beta_0, & \mu_1=\mu_0\\
\beta_2=\beta_0, & \mu_2=\mu_0
\end{array}\right.}
+ (\mbox{corrections of order}\ge 2).
\end{equation}
Therefore as far as we consider $J^{(E)}_{j,mn}$ up to the first order in
$\delta\beta$ and $\delta\mu$ (linear transport regime) we can replace
$\rho_{mm}$ in the definition (\ref{currentA}) into
\begin{equation}
\tilde{\rho}_{mm}={\rho_{mm}}_{\small\left|
\begin{array}{cc}
\beta_1=\beta_0, & \mu_1=\mu_0\\
\beta_2=\beta_0, & \mu_2=\mu_0
\end{array}\right.}
\end{equation}
Using
\begin{subequations}
\begin{eqnarray}
{\rm Re}\gamma_{-,1,\epsilon_m-\epsilon_n}-{\rm Re}\gamma_{-,2,\epsilon_m-\epsilon_n}&=&\gamma_{mn}\left(\delta\beta\frac{\partial}{\partial \beta_0}-\delta\mu\frac{\partial}{\partial \mu_0}\right)
\frac{e^{(\epsilon_m-\epsilon_n-\mu_0)\beta_0}}{e^{(\epsilon_m-\epsilon_n-\mu_0)\beta_0}-1}+\mbox{higher order correction}\\
{\rm Re}\gamma_{+,1,\epsilon_m-\epsilon_n}-{\rm Re}\gamma_{+,2,\epsilon_m-\epsilon_n}&=&\gamma_{mn}\left(\delta\beta\frac{\partial}{\partial \beta_0}-\delta\mu\frac{\partial}{\partial \mu_0}\right)
\frac{1}{e^{(\epsilon_m-\epsilon_n-\mu_0)\beta_0}-1}+\mbox{higher order correction}\\
\gamma_{mn}&=&\pi\int_{k\in\Omega_{mn}}dk~|g_{0,\epsilon_m-\epsilon_n}(k)|^2
\delta(\omega(k)-(\epsilon_m-\epsilon_n))
\end{eqnarray}
\end{subequations}
we get (we denote the approximate currents $\tilde{J}^{(E)}_{j,mn}$)
\begin{eqnarray}\label{J1-2}
\tilde{J}_{2\rightarrow1,mn}&:=&\frac{1}{2}
\left(\tilde{J}_{1,mn}-\tilde{J}_{2,mn}\right)
\nonumber\\
&=&\gamma_{mn}\left(\tilde{\rho}_{mm}
\left(\delta\beta\frac{\partial}{\partial \beta_0}-\delta\mu\frac{\partial}{\partial \mu_0}\right)
\frac{e^{(\epsilon_m-\epsilon_n-\mu_0)\beta_0}}{e^{(\epsilon_m-\epsilon_n-\mu_0)\beta_0}-1}-\tilde{\rho}_{nn}
\left(\delta\beta\frac{\partial}{\partial \beta_0}-\delta\mu\frac{\partial}{\partial \mu_0}\right)
\frac{1}{e^{(\epsilon_m-\epsilon_n-\mu_0)\beta_0}-1}\right)
\nonumber\\
&=&\frac{\gamma_{mn}}{\tilde{Z}}\left[\tilde{\rho}_{mm}
\left(\frac{\delta\beta}{\beta_0}(\epsilon_m-\epsilon_n-\mu_0)
-\delta\mu\right)\frac{\partial}{\partial(\epsilon_m-\epsilon_n)}
\frac{e^{(\epsilon_m-\epsilon_n-\mu_0)\beta_0}}{e^{(\epsilon_m-\epsilon_n-\mu_0)\beta_0}-1}\right.\nonumber\\
&&
\hspace*{2cm}
\left.
-\tilde{\rho}_{nn}
\left(\frac{\delta\beta}{\beta_0}(\epsilon_m-\epsilon_n-\mu_0)
-\delta\mu\right)\frac{\partial}{\partial(\epsilon_m-\epsilon_n)}
\frac{1}{e^{(\epsilon_m-\epsilon_n-\mu_0)\beta_0}-1}\right]
\\
J^{E}_{2\rightarrow 1,mn}&=&(\epsilon_m-\epsilon_n) J_{2\rightarrow 1,mn}
\end{eqnarray}
In addition, if $\epsilon_m \gg \mu_0 \gg \delta\mu$, 
by the equilibrium approximation
\begin{subequations}\label{yokei_kinji}
\begin{equation}
\tilde{\rho}_{mm}=\frac{1}{\tilde Z}e^{-\beta_0 \epsilon_m},\quad
\tilde{Z}=\sum_{m}e^{-\beta_0 \epsilon_m},
\end{equation}
one can see
\begin{equation}\label{linear_current}
\tilde{J}^{(E)}_{2\rightarrow 1,mn}=\tilde{J}^{(E)}_{1,mn}=
-\tilde{J}^{(E)}_{2,mn}\neq 0,
\end{equation}
\end{subequations}
which hold the condition (\ref{dynamical_db_condition}).

From this it is clear that, for the system $S$ (say atom),
the non-equilibrium effects appear as {\it first order} effects in
the currents (\ref{linear_current}), but only as {\it second order} terms
in the state. This suggests a theoretical explanation of
both the empirical success and the limitations of Kubo linear response
theory.

Let us show a relation \cite{Onsager31} between the two currents
$\tilde{J}_{1\rightarrow 2,mn}$ and 
$\tilde{J}^Q_{1\rightarrow 2,mn}
=\tilde{J}^E_{1\rightarrow 2,mn}-\mu_0\tilde{J}_{1\rightarrow 2,mn}$,
which is the analogue of the famous Onsager relation between the electric and
heat currents in the conductivity problem. It is only
an analogy because the carrier of our currents 
is a Boson particle and  not Fermion (electron).
From (\ref{J1-2}), 
\begin{subequations}\label{onsager}
\begin{equation}\left[
\begin{array}{c}
\tilde{J}_{1\rightarrow 2,mn}\\
\tilde{J}^Q_{1\rightarrow 2,mn}
\end{array}
\right]
=\left[
\begin{array}{cc}
\Gamma_{mn} & L_{mn}\\
L_{mn} & M_{mn}
\end{array}
\right]\left[
\begin{array}{c}
\delta\mu_0 \\
\delta\beta\over\beta_0
\end{array}\right]
\end{equation}
where
\begin{eqnarray}
\Gamma_{mn}&=&-{\gamma}_{mn}
\left(\tilde{\rho}_{mm}
\frac{\partial}{\partial(\epsilon_m-\epsilon_n)}
\frac{e^{(\epsilon_m-\epsilon_n-\mu_0)\beta_0}}{e^{(\epsilon_m-\epsilon_n-\mu_0)\beta_0}-1}
-
\tilde{\rho}_{nn}
\frac{\partial}{\partial(\epsilon_m-\epsilon_n)}
\frac{1}{e^{(\epsilon_m-\epsilon_n-\mu_0)\beta_0}-1}
\right) \nonumber\\
&=&\frac{\gamma_{mn}\beta_0e^{(\epsilon_m-\epsilon_n-\mu_0)\beta_0}}
{(e^{(\epsilon_m-\epsilon_n-\mu_0)\beta_0}-1)^2}
\left(\tilde{\rho}_{mm}-\tilde{\rho}_{nn}\right) ~~(<0 \quad 
\mbox{when (\ref{state_futuu}) holds.})
\\
L_{mn}&=&-(\epsilon_m-\epsilon_n-\mu_0)\Gamma_{mn},\quad
M_{mn}=-(\epsilon_m-\epsilon_n+\mu_0)^2\Gamma_{mn}
\end{eqnarray}
\end{subequations}
or we obtain explicitly
\begin{equation}\label{coooooool}
\frac{\partial \tilde{J}_{1\rightarrow 2,mn}}
{\partial \left(\frac{\delta\beta}{\beta_0}\right)}
=\frac{\partial \tilde{J}^Q_{1\rightarrow 2,mn}}
{\partial \delta\mu}=L_{mn},
\end{equation}
which is the Onsager reciprocal relation.

One can easily see that these currents produce positive entropy.
Following \cite{Prigogine62}, 
the entropy production with these currents is given as
\begin{eqnarray}\label{entropy_production}
\sigma:&=&
\beta_0\left(\tilde{J}_{1\rightarrow 2,mn}(-\delta\mu)+\tilde{J}^{Q}_{1\rightarrow 2,mn}
\frac{\delta\beta}{\beta_0}\right)\nonumber\\
&=&\beta_0 \left(-\Gamma_{mn} \delta\mu^2-2 L_{mn}\delta\mu\frac{\delta\beta}{\beta_0}+M_{mn}\left(\frac{\delta\beta}{\beta_0}\right)^2\right),
\end{eqnarray}
and as far as (\ref{state_futuu}) holds, since $L^2_{mn}+\Gamma_{mn}M_{mn}=0$,
$\delta S$ is positive for any $(\delta\mu,\delta\beta)$ except for
\begin{equation}\label{minimum_condition}
\delta\mu=\delta\beta=0,\quad \delta\mu=(\epsilon_m-\epsilon_n-\mu_0)
\frac{\delta\beta}{\beta_0}
\end{equation}
which imply $\tilde{J}^{(E)}_{1\rightarrow 2,mn}=0$.

As is well known, Onsager reciprocal relation
is understood as a consequence of microscopic symmetry of the dynamics, 
based on  the following two assumptions \cite{Onsager31}:
(i) There exists an intermediate time scale between macro and micro dynamics.
(ii) Average of spontaneous thermal fluctuation of the microscopic
observable decaying is described by macroscopic transport theory.
Notice that both the above assumptions were deduced in our model
from the stochastic limit. (i) corresponds to the fact that
the convergence to the stationary state of the system is described in
the rescaled time scale. This time scale is exactly the time scale
used in assumption (i).
Moreover what the stochastic limit tells us is that the dynamics
of the currents (or the transport coefficients) are given in terms of 
the time correlations of the original field in the initial
state. This is nothing but the situation described by assumption (ii).
In the context of derivation of the Onsager relation between heat and 
electric currents by linear response theory, since there is no
Hamiltonian which can describe the force generating a heat current
whereas chemical potential can be treated always dynamically,
(ii) has to be required as assumption\cite{Kubo57}.
In the present paper, both temperature and chemical potential are treated
as parameters of the environment fields in the framework of the quantum
mechanics for a open system. 
Moreover one should notice that the current is described directly
in terms of the dynamics of the fields.
It is also important to notice that the equilibrium state
approximation (\ref{yokei_kinji}) is not necessary to derive the Onsager
relation (\ref{coooooool}). Usually, Onsager relation is
derived assuming a symmetric property of the microscopic dynamics
\cite{Onsager31}.
However as is discussed in the next section,
this symmetric property is equivalent to the requirement that
the state is equilibrium (see below (\ref{eq:on_symm})).
Our results prove that the Onsager reciprocal relation (\ref{coooooool})
can be valid without any symmetry of the dynamics.
Gabrielli, Jona-Lasinio and Landim illustrated such a possibility using
a classical, solvable and phenomenological model~\cite{Jona96}.

\section{Master equation characterization of dynamical detailed balance}
\label{time symmetricity}
In the equilibrium case, it is well known that the detailed balance
condition can be characterized by a generator of the master equation of
the system interacting with the environment \cite{Gorini-Kossokowski76,Kossakowski-Frigerio-Gorini-Verri77}.
Given the dynamical semigroup which drives the state to an equilibrium state
\begin{equation}
\frac{d}{dt}\rho_t={\cal L}^*\rho_t,\quad \rho_t\rightarrow \rho_{eq},
\end{equation}
where
\begin{equation}\label{L*}
{\rm tr}\left(X{\cal L}^*(\rho_t)\right)
={\rm tr}\left(\rho_t {\cal L}(X)\right).
\end{equation}
The detailed balance condition or KMS condition
for $\rho_{eq}$ is characterized by the following equations~\cite{Kossakowski-Frigerio-Gorini-Verri77}:
\begin{subequations}\label{dyn_ch_eq}
\begin{equation}\label{L+}
{\rm tr}\left(\rho_{eq}{\cal L}^+(A)B\right):=
{\rm tr}\left(\rho_{eq}A{\cal L}(B)\right)
\quad\text{for all }~A, B
\end{equation}
\begin{equation}\label{Lsymmetry}
{\cal L}(X)-{\cal L}^+(X)=2i[H,X]\quad (H=H^\dagger) \quad\text{for all }~X
\end{equation}
\end{subequations}
In this section, we prove a generalization of the above characterization to
non-equilibrium stationary states in terms of the dynamical detailed
balance condition defined in the previous section.

We consider the forward and the backward Heisenberg evolution of
a system operator $X$, i.e. (cf. \cite{AcLuVo} Chap I, section 1.1.29)
\begin{equation}
j^{(F)}_t(X):=U^\dagger_t X U_t~~\text{for}~~t>0,\qquad
j^{(B)}_t(X):=U_{-t} X U^\dagger_{-t}~~\text{for}~~t<0
\end{equation}
where $U_t$ is the time evolution operator in interaction picture. After
stochastic limit and in the notations (17), (23), these lead to the master equations
for observables
\begin{subequations}
\label{eq:heisenberg}
\begin{eqnarray}
\frac{d}{dt}\langle j^{(F)}_t(X) \rangle=&&
i[\Delta, \langle j_t^{(F)}\rangle]\nonumber\\
&&-\sum_{\omega\in F}\left(
\Gamma_{\omega-}\left(\frac{1}{2}\{E_\omega^\dagger E_\omega, 
\langle j_t^{(F)}(X)\rangle \}-E_\omega^\dagger \langle j_t^{(F)}(X)\rangle E_\omega\right)\right.\nonumber\\
&&\left.\hspace{1cm}
+\Gamma_{\omega+}\left(\frac{1}{2}\{E_\omega E_\omega^\dagger, 
\langle j_t^{(F)}(X)\rangle \}-E_\omega \langle j_t^{(F)}(X)\rangle E_\omega^\dagger\right)\right)\nonumber\\
=:&&{\cal L}(\langle j^{(F)}_t(X) \rangle),\quad \text{for}~~t \ge 0
\label{forward}
\end{eqnarray}
\begin{eqnarray}
\frac{d}{dt}\langle j^{(B)}_t(X) \rangle=&&
i[\Delta, \langle j_t^{(B)}\rangle]\nonumber\\
&&+\sum_{\omega\in F}
\left(\Gamma_{\omega-}\left(\frac{1}{2}\{E_\omega^\dagger E_\omega, 
\langle j_t^{(B)}(X)\rangle \}-E_\omega^\dagger \langle j_t^{(B)}(X)\rangle E_\omega\right)\right.\nonumber\\
&&\left.\hspace{1cm}
+\Gamma_{\omega+}\left(\frac{1}{2}\{E_\omega E_\omega^\dagger, 
\langle j_t^{(B)}(X)\rangle \}-E_\omega \langle j_t^{(B)}(X)\rangle E_\omega^\dagger\right)\right)\nonumber\\
=:&&
-{\cal L}_{B}(\langle j^{(B)}_t(X) \rangle),\quad \text{for}~~t \le 0.
\label{backward}
\end{eqnarray}
\end{subequations}
where $\langle\cdot\rangle$ denotes partial trace of the field degrees of freedom.
Through (\ref{L*}), the dual master equation (\ref{eq:master_equation}) (for density matrices)
is written as
\begin{equation}\label{master_forward}
\frac{d}{dt}\rho_S(t)={\cal L}^*\rho_S(t),\quad t\ge 0.
\end{equation}
Similarly, we introduce a master equation associated to ${\cal L}_{B}$ as
\begin{equation}\label{master_backward}
\frac{d}{dt}\rho^{(B)}_S(t)=-{\cal L}_{B}^*\rho^{(B)}_S(t),\quad
t \le 0.
\end{equation}
Both master equations have the
same stationary state $\rho_S$ (see (\ref{eq:offdiagonal_equation}) and
(\ref{eq:diagonal_equation})).

As easily seen from (\ref{eq:heisenberg}), with $\Delta=\Delta^\dagger$ given by (23) one has
\begin{subequations}\label{eq:daiji}
\begin{equation}\label{eq:itumo}
{\cal L}(X)-{\cal L}_B(X)=2i[\Delta,X].
\end{equation}
By direct computation we obtain the deviation from the symmetry condition 
${\rm tr}(\rho_{S}x{\cal L}(y))=
{\rm tr}(\rho_{S}{\cal L}_B(x)y)$ which characterizes equilibrium:
\begin{eqnarray}\label{eq:daiji_1}
&&{\rm tr}\left(\rho_{S}X{\cal L}(Y)\right)-{\rm tr}\left(\rho_{S}{\cal L}_B(X)Y\right)=
\sum_{lm}X_{ll}Y_{mm}
\left(\rho_{ll}
(\Gamma_{-,\epsilon_l-\epsilon_m}+\Gamma_{+,\epsilon_m-\epsilon_l})
-\rho_{mm}(\Gamma_{-,\epsilon_m-\epsilon_l}+\Gamma_{+,\epsilon_l-\epsilon_m})\right)\nonumber\\
&&=\sum_{lm}X_{ll}Y_{mm}
\theta(\epsilon_l-\epsilon_m)(J_{1,lm}+J_{2,lm})-
\theta(\epsilon_m-\epsilon_l)(J_{1,ml}+J_{2,ml}) 
\end{eqnarray}
where
\begin{equation}
X_{ll}=\langle \epsilon_l|X|\epsilon_l\rangle,\quad
Y_{mm}=\langle \epsilon_m|Y|\epsilon_m\rangle,\quad
\rho_{ll}=\langle \epsilon_l|\rho_{S}|\epsilon_l\rangle.
\end{equation}
\end{subequations}
Choosing
\begin{equation}
X=|\epsilon_a\rangle\langle\epsilon_a|=:P_a,\quad 
Y=|\epsilon_b\rangle\langle\epsilon_b|=:P_b,
\end{equation}
(\ref{eq:daiji_1}) becomes
\begin{equation}\label{eq:daiji_sp}
{\rm tr}\left(\rho_{S}P_a{\cal L}(P_b)\right)-{\rm tr}\left(\rho_{S}{\cal L}_B(P_a)P_b\right)=
\theta(\epsilon_a-\epsilon_b)(J_{1,ab}+J_{2,ab})-
\theta(\epsilon_b-\epsilon_a)(J_{1,ba}+J_{2,ba}).
\end{equation}
The left hand side describes
the balance between two processes:
transition from $|\epsilon_a\rangle$ to $|\epsilon_b\rangle$ and its converse
in stationary state $\rho_{S}$. Thus (\ref{eq:daiji}) (or (\ref{eq:daiji_sp}))
is a characterization of the dynamical detailed balance condition discussed
the previous section. Remember usual detailed balance condition is
characterized by (\ref{dyn_ch_eq}) which is the case when the right hand side
of (\ref{eq:daiji_1}) is identically zero.

Notice that $\rho_{S}$ is an equilibrium 
state when $J_{1,mn}+J_{2,mn}=0$. Let us remark again that
as far as linear approximation is
concerned,
$\tilde{J}_{1,mn}=\tilde{J}_{2,mn}=0$ is not necessary to realize an
equilibrium state $\tilde{\rho}_{eq}$
(the equilibrium approximation (\ref{yokei_kinji}))
which follows the condition 
(\ref{dyn_ch_eq}) up to the first order (see Sec.\ref{subSec.linear_app}).
In this case,
\begin{eqnarray}\label{eq:on_symm}
{\rm tr}\left(\tilde{\rho}_{eq}X{\cal L}(Y)\right)
-{\rm tr}\left(\tilde{\rho}_{eq}{\cal L}_B(X)Y\right)&=&
{\rm tr}\left(\tilde{\rho}_{eq}X{\cal L}(Y)\right)
-{\rm tr}\left(\tilde{\rho}_{eq}{\cal L}(X)Y\right)\nonumber\\
&=&0
\end{eqnarray}
and it is exactly the symmetry
of microscopic dynamics assumed in the original derivation of
Onsager law\cite{Onsager31}.

\section{Local KMS condition}
The KMS condition is known to be a 
characterization of equilibrium states equivalent to the
detailed balance condition.
In this section, we prove that a generalization of the KMS condition which
characterizes the state described with the dynamical detailed balance
condition. 

First, we introduce a generalization of the KMS condition which distinguishes 
between
those general density matrices which commutes with a given discrete
Hamiltonian and those which are function of the given Hamiltonian.
This condition, which we call {\it local KMS condition} in
the sense of energy space, can describe states with
mode-dependent temperatures

Given a discrete spectrum  Hamiltonian $H_S$:
\begin{equation}
H_S=\sum_{\epsilon}\epsilon P_{\epsilon}\quad,\quad
P_{\epsilon}=|\epsilon\rangle\langle\epsilon| \quad,\quad
H_S|\epsilon\rangle=\epsilon|\epsilon\rangle
\end{equation}
For any complex valued Borel function $f :\mathbb{R}\rightarrow \mathbb{C}$
the map $x\mapsto e^{itf(H_S)}xe^{-itf(H_S)}$ is defined by the spectral
theorem and one has
\begin{subequations}
\begin{equation}\label{meaning}
x(t):=e^{itf(H_S)}xe^{-itf(H_S)}
=\sum_{\epsilon,\epsilon'}e^{it(f(\epsilon)-f(\epsilon'))}
P_{\epsilon}xP_{\epsilon'}=
\sum_{\delta\in B_f} e^{it\delta} E_{\delta}^{f}(x)
\end{equation}
where
\begin{equation}
B_f:=\{f(\epsilon)-f(\epsilon') ;~\forall \epsilon, \epsilon'\}\quad ,
\quad E_{\delta}^{f}(x) :=
\sum_{\epsilon, \epsilon' \ : \ f(\epsilon)-f(\epsilon') = \delta}
P_{\epsilon}xP_{\epsilon'}.
\end{equation}
\end{subequations}
For such Hamiltonian $H_S$ the following theorem holds:\\
\begin{theorem}
\begin{subequations}\label{GKMS}
For a density matrix $\rho$ and the corresponding state
$\langle\langle \ \cdot \ \rangle\rangle$ the following are equivalent:\\
(i) There exists a real valued Borel function
$\beta :{\mathbb{R}}\rightarrow {\mathbb{R}}$ 
such that $\exp-\beta(H_S) H_S$ is
trace class and
\begin{equation}\label{eq:state}
\rho=\frac{1}{Z}e^{-\beta(H_S) H_S}
\end{equation}
(ii) There exists a real valued Borel function
$\beta :\mathbb{R}\rightarrow \mathbb{R}$ such that $\exp-\beta(H_S) H_S$ is
trace class and $\rho$ satisfies the following local KMS condition with
respect to the Heisenberg dynamics $x \mapsto e^{itH_S}xe^{-itH_S}$:
\begin{equation}\label{eq:LKMS}
\forall x,y,t,
\quad \langle\langle xy(t+i\beta(H_S))\rangle\rangle
=\langle\langle y(t) x \rangle\rangle
\end{equation}
where the meaning of $y(t+i\beta(H_S))$ is given by (\ref{meaning}).
\end{subequations}
\end{theorem}
\noindent
{\it Proof}. \\
{\bf (\ref{eq:state}) $\Rightarrow$ (\ref{eq:LKMS}).}
\begin{equation}
\langle\langle x y(t+i\beta(H_S)) \rangle =
{\rm tr}\left(\rho x e^{-\beta(H_S)H_S}y(t)e^{+\beta(H_S)H_S}\right)
= \frac{1}{Z}{\rm tr}\left(x e^{-\beta(H_S)H_S}y(t)\right)
= {\rm tr}\left(y(t)x \rho\right) = \langle\langle y(t) x\rangle\rangle
\end{equation}
{\bf (\ref{eq:LKMS}) $\Rightarrow$ (\ref{eq:state}).}
\\(\ref{eq:LKMS}) means
that for all  $x,y$ and for all $t$
\begin{equation}
{\rm tr}\left(e^{-\beta(H_S)H_S}y(t)e^{+\beta(H_S)H_S}\rho x\right)
={\rm tr} \left(\rho y(t)x\right)
\end{equation}
Therefore for all $y$ and for all $t$
\begin{equation}
e^{-\beta(H_S)H_S}y(t)e^{+\beta(H_S)H_S}\rho=\rho y(t)
\end{equation}
or equivalently, putting $t=0$ and replacing $y$ by $ye^{-\beta(H_S)H_S}$
\begin{equation}\label{11}
e^{-\beta(H_S)H_S}y \rho=\rho e^{\beta(H_S)H_S}y
\end{equation}
hence, putting $y=1$
\begin{equation}\label{12}
e^{\beta(H_S)H_S}\rho=\rho e^{\beta(H_S)H_S}
\end{equation}
(\ref{11}),(\ref{12}) imply that, for all $y$
\begin{equation}
y e^{\beta(H_S)H_S}\rho=e^{\beta(H_S)H_S}\rho y
\end{equation}
and this implies that, for some scalar $\lambda$
\begin{equation}
\label{13}
e^{\beta(H_S)H_S}\rho=\lambda 1
\end{equation}
Since ${\rm tr}(\rho)=1$, (\ref{13}) implies that
\begin{equation}
\rho=\frac{1}{Z} e^{-\beta(H_S)H_S}.
\end{equation}
{\it (Q.E.D)}

Notice that when $\beta(H_S)=\beta$ (constant),
the state (\ref{eq:state}) is the Gibbs state
at temperature $\beta^{-1}$ and (\ref{eq:LKMS}) becomes the KMS condition.

We shall prove that this local KMS condition (\ref{GKMS}) is equivalent 
to the dynamical detailed balance condition (\ref{eq:daiji}).
To avoid infinite-valued functions, we assume that all the $\rho_{ll}$ are
strictly positive and we represent the stationary solution $\rho_{S}$ of
(\ref{master_forward}) and (\ref{master_backward}) in the form
\begin{equation}
\rho_{S}=\frac{1}{Z}e^{-\beta_S(H_S)H_S},\quad \beta_{S}(\epsilon_l)=
-\frac{1}{\epsilon_l}\log \rho_{ll}.
\end{equation}
For such state the following theorem holds:

\begin{theorem}
The dynamical detailed balance condition (\ref{eq:daiji}) holds
if and only if the local KMS condition (\ref{GKMS}) is
satisfied.
\end{theorem}

\noindent
{\it Proof}. {\bf (\ref{GKMS}) $\Rightarrow$ (\ref{eq:daiji}).}

Appling the local KMS-condition (\ref{GKMS}) to this state,
we get
\begin{equation}\label{proof_sentou}
\langle\langle AB \rangle\rangle=\langle\langle B(-i\beta_S(H_S))A
\rangle\rangle,
\end{equation}
In addition in the notations (17), (23) and using relations
\begin{subequations}
\begin{equation}
\Delta(-i\beta_S(H_S))=\Delta
\end{equation}
\begin{equation}
E_{\epsilon_m-\epsilon_n}(-i\beta_S(H_S))=
e^{\beta_S(\epsilon_n)\epsilon_n-\beta_S(\epsilon_m)\epsilon_m}
E_{\epsilon_m-\epsilon_n},\quad
E^\dagger_{\epsilon_m-\epsilon_n}(-i\beta_S(H_S))=
e^{\beta_S(\epsilon_m)\epsilon_m-\beta_S(\epsilon_n)\epsilon_n}
E_{\epsilon_m-\epsilon_n}^\dagger,
\end{equation}
\end{subequations}
we obtain
\begin{subequations}\label{eq:85}
\begin{eqnarray}
\langle\langle X [\Delta,Y]\rangle\rangle&=&
\langle\langle X\Delta Y-\Delta(-i\beta_S(H_S))XY \rangle\rangle\nonumber\\
&=&\langle\langle X\Delta Y-\Delta XY\rangle\rangle\nonumber\\
&=&\langle\langle[X,\Delta]Y\rangle\rangle\\
\langle\langle X \{E^\dagger_{\epsilon_m-\epsilon_n} E_{\epsilon_m-\epsilon_n},
Y\}\rangle\rangle
&=&\langle\langle X 
E^\dagger_{\epsilon_m-\epsilon_n} E_{\epsilon_m-\epsilon_n} Y
+XY E^\dagger_{\epsilon_m-\epsilon_n} E_{\epsilon_m-\epsilon_n}
\rangle\rangle\nonumber\\
&=&\langle\langle X 
E^\dagger_{\epsilon_m-\epsilon_n} E_{\epsilon_m-\epsilon_n} Y
+E^\dagger_{\epsilon_m-\epsilon_n}(-i\beta_S(H_S))
E_{\epsilon_m-\epsilon_n}(-i\beta_S(H_S))XY
\rangle\rangle\nonumber\\
&=&\langle\langle X 
E^\dagger_{\epsilon_m-\epsilon_n} E_{\epsilon_m-\epsilon_n} Y
+E^\dagger_{\epsilon_m-\epsilon_n} E_{\epsilon_m-\epsilon_n} 
XY\rangle\rangle\nonumber\\
&=&\langle\langle
\{X,E^\dagger_{\epsilon_m-\epsilon_n} E_{\epsilon_m-\epsilon_n}\}Y
\rangle\rangle\\
\langle\langle X E^\dagger_{\epsilon_m-\epsilon_n}YE_{\epsilon_m-\epsilon_n}
\rangle\rangle
&=&e^{\beta_S(\epsilon_n)\epsilon_n-\beta_S(\epsilon_m)\epsilon_m}
\langle\langle E_{\epsilon_m-\epsilon_n} X E^\dagger_{\epsilon_m-\epsilon_n}
Y\rangle\rangle\\
\langle\langle X E_{\epsilon_m-\epsilon_n}YE^\dagger_{\epsilon_m-\epsilon_n}
\rangle\rangle
&=&e^{\beta_S(\epsilon_m)\epsilon_m-\beta_S(\epsilon_n)\epsilon_n}
\langle\langle E^\dagger_{\epsilon_m-\epsilon_n} X E_{\epsilon_m-\epsilon_n}
Y\rangle\rangle.
\end{eqnarray}
\end{subequations}
Now let us define ${\cal L}^+_G$ by the relation:
\begin{equation}\label{eq L+G}
\langle\langle {\cal L}^{+}_G(X)Y\rangle\rangle:=
\langle\langle X{\cal L}(Y)\rangle\rangle
\end{equation}
for ${\cal L}$ given by (\ref{forward}). Notice that we are defining
${\cal L}^{+}_G$ not only in equilibrium state but also in the non-equilibrium
stationary state which is described with the local KMS condition
(\ref{GKMS}), unlike (\ref{L+}).
Using relation (\ref{eq:85}), we find 
$(\omega=\epsilon_m-\epsilon_n)$
\begin{subequations}\label{eq:keisanG}
\begin{eqnarray}
{\cal L}^+_G(X)&=&-i[\Delta,X]-\sum_{\omega\in F}\left(
\Gamma_{-,\omega}
\left(\frac{1}{2}\{E^\dagger_\omega E_\omega,X\}-E^\dagger_\omega X E_\omega\right)
+\Gamma_{+,\omega}
\left(\frac{1}{2}\{E_\omega E^\dagger_\omega,X\}-E_\omega X E^\dagger_\omega\right)\right)\nonumber\\
&&+\sum_{\omega\in F}\left((\Gamma_{+,\omega}
e^{\beta_S(\epsilon_m)\epsilon_m-\beta_S(\epsilon_n)\epsilon_n}-\Gamma_{-,\omega})E^\dagger_\omega X E_\omega
+
(\Gamma_{-,\omega}
e^{\beta_S(\epsilon_n)\epsilon_n-\beta_S(\epsilon_m)\epsilon_m}-\Gamma_{+,\omega})E_\omega X E^\dagger_\omega\right)\nonumber\\
&=&{\cal L}_{B}(X)+\sum_{\omega\in F}\hat{\Pi}_{\omega}(X)\\
\hat{\Pi}_{\omega}(X)&=&(\Gamma_{+,\omega}
e^{\beta_S(\epsilon_m)\epsilon_m-\beta_S(\epsilon_n)\epsilon_n}-\Gamma_{-,\omega})E^\dagger_\omega X E_\omega
+(\Gamma_{-,\omega}
e^{\beta_S(\epsilon_n)\epsilon_n-\beta_S(\epsilon_m)\epsilon_m}-\Gamma_{+,\omega})E_\omega X E^\dagger_\omega
\end{eqnarray}
\end{subequations}
(\ref{eq L+G}) and (\ref{eq:keisanG}) mean
\begin{equation}\label{eq:88}
\langle\langle X{\cal L}(Y)\rangle\rangle=
\langle\langle{\cal L}_B(X) Y
\rangle\rangle+\sum_{\omega\in F}\langle\langle\hat{\Pi}_{\omega}(X)Y
\rangle\rangle
\end{equation}
and
\begin{eqnarray}\label{eq:89}
\sum_{\omega\in F}\langle\langle\hat{\Pi}_{\omega}(X)Y
\rangle\rangle
&=&\sum_{\omega\in F}{\rm tr}\left(\frac{e^{-\beta_S(H_S)H_S}}{Z}
\left((\Gamma_{+,\omega}
e^{\beta_S(\epsilon_m)\epsilon_m-\beta_S(\epsilon_n)\epsilon_n}-\Gamma_{-,\omega})E^\dagger_\omega X E_\omega\right.\right.\nonumber\\
&&\left.\left.
\hspace{3cm}+(\Gamma_{-,\omega}
e^{\beta_S(\epsilon_n)\epsilon_n-\beta_S(\epsilon_m)\epsilon_m}-\Gamma_{+,\omega})E_\omega X E^\dagger_\omega\right)Y\right)
\nonumber\\
&=&\sum_{\epsilon_m,\epsilon_n}
\left(X_{nn}Y_{mm}\left(\Gamma_{+,\epsilon_m-\epsilon_n}\rho_{nn}-
\Gamma_{-,\epsilon_m-\epsilon_n}\rho_{mm}\right)\right.\nonumber\\
&&\hspace{3cm}\left.
+X_{mm}Y_{nn}
\left(
\Gamma_{-,\epsilon_m-\epsilon_n}\rho_{mm}-\Gamma_{+,\epsilon_m-\epsilon_n}
\rho_{nn}\right)\right)\nonumber\\
&=&\sum_{\epsilon_m,\epsilon_n}X_{nn}Y_{mm}
\left(\Gamma_{+,\epsilon_m-\epsilon_n}\rho_{nn}-
\Gamma_{-,\epsilon_m-\epsilon_n}\rho_{mm}+
\Gamma_{-,\epsilon_n-\epsilon_m}\rho_{nn}-
\Gamma_{+,\epsilon_n-\epsilon_m}\rho_{mm}\right)\nonumber\\
&=&\sum_{\epsilon_m,\epsilon_n}X_{nn}Y_{mm}
\theta(\epsilon_n-\epsilon_m)(J_{1,nm}+J_{2,nm})-
\theta(\epsilon_m-\epsilon_n)(J_{1,mn}+J_{2,mn}) 
\end{eqnarray}
(\ref{eq:88}) and (\ref{eq:89}) is exactly the dynamical detailed balance
condition (\ref{eq:daiji}). 

\noindent
{\bf (\ref{eq:daiji}) $\Rightarrow$ (\ref{GKMS}).}

Following (\ref{proof_sentou})$\sim$(\ref{eq:89}) conversely,
we see that the dynamical detailed balance condition (\ref{eq:daiji}) implies
\begin{equation}
{\rm tr}\left(\rho_{S}X{\cal L}(Y)\right)
={\rm tr}\left(\rho_{S} e^{\beta_S(H_S)H_S}{\cal L}(Y)e^{-\beta_S(H_S)H_S}X
\right),\quad \forall ~X,Y.
\end{equation}
For off diagonal type operator
$$
\tilde{y}=\sum_{m\neq n}C_{mn}|\epsilon_m\rangle\langle \epsilon_n|
$$
there exists $Y$ such that
\begin{equation}
{\cal L}(Y)=\tilde{y}
\end{equation}
and putting $\tilde{y}=e^{itH_S}ye^{-itH_S}$ ($y$ is also off diagonal type)
we get
$$
{\rm tr}\left(\rho_{S}X e^{itH_S}ye^{-itH_S} \right)
={\rm tr}\left(\rho_{S} e^{\beta_S(H_S)H_S}e^{itH_S}
ye^{-itH_S}e^{-\beta_S(H_S)H_S}X
\right),\quad \forall ~X.
$$
or
\begin{equation}\label{pr:off_diagonal}
{\rm tr}\left(\rho_{S}X y(t) \right)=
{\rm tr}\left(\rho_{S} y(t+i\beta_S(H_S))X \right),\quad \forall ~X.
\end{equation}
In addition, since $\rho_{st}$ is diagonal, (\ref{pr:off_diagonal}) is always
satisfied with any diagonal type operator
$y=\sum_{m}C_{mm}|\epsilon_m\rangle\langle\epsilon_m|$ also.
Therefore, (\ref{pr:off_diagonal}) is always satisfied with any operator $X$
and $y$. \\
\noindent{\it (Q.E.D)}

Notice that since 
$${\cal L}_G^+(1)=\sum_{\omega}\hat{\Pi}_{\omega}(1)\neq 0$$ in
the non-equilibrium case, ${\cal L}_G^+$ cannot be a generator of
any dynamical semigroup whereas ${\cal L}_{B}$  always exists
as generator of dynamical semigroup.
This is also one of the particular properties of the
non-equilibrium state.
In an equilibrium case, as we have seen $\beta_S(x)$ become a constant $\beta$
which is the same inverse temperature of the environment fields, and the
equality
${\Gamma_{+,\omega}}/{\Gamma_{-,\omega}}=e^{-\beta\cdot(\epsilon_m-\epsilon_n)}$holds, i.e. $\hat{\Pi}_\omega(X)=0$ which implies ${\cal L}_B={\cal L}_G^+$.

\section{Interaction with non-equilibrium field}
\label{ModelB}
In the previous sections, we considered the non-equilibrium stationary
states of a system driven by two environments at two different
temperatures and we discussed several characterizations of such states.
In this section, applying these characterizations to the state of the
environment,
we consider a system interacting with an environment in local equilibrium.
(On the local KMS condition for the field degrees of freedom, see the next
section.) 
One will see not only that the stationary state of the system driven by such
non-equilibrium environment can be characterized as for the previous
model, but also that interesting non-linear effects due to
the interaction with non-equilibrium environment exist
whose physical meaning is different from the previous model.

We consider a system interacting with a single boson field whose state is
described by a generalized temperature function.
Technically, the analysis of the model can be done in the same way as
the previous one.
Instead of (\ref{eq:Hamiltonian1}) but similarly, the Hamiltonian
\begin{subequations}
\begin{equation}
H=H_0+\lambda H_I,\quad (\lambda ~\mbox{is a coupling constant.})
\end{equation}
\begin{equation}
H_0=H_S+H_B,\quad
H_S=\sum_{l}\epsilon_l |\epsilon_l\rangle\langle\epsilon_l|,\quad
H_B=\int \omega(k)a^\dagger_{k}a_{k}
\quad [a_{k},a_{k'}^\dagger]=\delta(k-k'), 
\end{equation}
\begin{equation}
H_I=\int dk~\left(g(k)Da^\dagger_{k}+g^*(k)D^\dagger a_{k}\right).
\end{equation}
\end{subequations}
On the other hand, we assume that the initial state of the field 
is a mean zero gauge invariant Gaussian state with correlations:
\begin{equation}\label{eq:initial_state2}
\langle a_{k}^\dagger a_{k'} \rangle=
N(k)\delta(k-k'),\quad
N(k)=\frac{1}{e^{\beta(\omega(k))\omega(k)}-1}
\end{equation}
where $\beta(\omega(k))$ is some positive function. This is a natural
generalization of the Gibbs factor to which it reduces when $\beta(\omega)$
is constant:
\begin{equation}\label{eq:constant_temperature}
\beta(\omega)=\beta.
\end{equation}
Exactly in the same way as in the previous argument, one can derive
the white noise
Hamiltonian equation
\begin{equation}\label{eq:white_noise_eq2}
\frac{d}{dt}U_t=-i \sum_{\omega\in F}
\left(E_\omega b^\dagger_{t:\omega}+E^\dagger_\omega 
b_{t:\omega}
\right)U_t
\end{equation}
where
\begin{equation}
b_{t:\omega}=
\int dk~g^*_{\omega}(k)b_{t:\omega}(k),\quad
b_{t:\omega}(k)=\lim_{\lambda\rightarrow 0}\frac{1}{\lambda}
e^{-i(\omega_j(k)-\omega)t/\lambda^2}a_{k}.
\end{equation}
The state of the limit white nose will be of the same type
with correlations
\begin{subequations}
\begin{eqnarray}
\langle b^\dagger_{t:\omega}(k) b_{t':\omega'}(k')\rangle
&=&\delta_{\omega\omega'}2\pi\delta(t-t')\delta(k-k')\delta(\omega(k)-\omega)N(k)\\
\langle b_{t:\omega}(k) b^\dagger_{t':\omega'}(k')\rangle
&=&\delta_{\omega\omega'}2\pi\delta(t-t')\delta(k-k')\delta(\omega(k)-\omega)\left(N(k)+1\right).
\end{eqnarray}
\end{subequations}
Finally we obtain the master equation (\ref{eq:master_equation}) but with 
different parameters
\begin{subequations}
\begin{eqnarray}
&&\Delta=i~\sum_{\omega\in F}\left({\rm Im}(\gamma_{-\omega})
E^\dagger_\omega E_\omega-{\rm Im}(\gamma_{+\omega})E_\omega E_\omega^\dagger\right)\\
&&\Gamma_{\mp,\omega}=2{\rm Re}~
\gamma_{\mp,\omega} \ge 0,\quad (\Gamma_{\mp,\omega}=0 \quad \mbox{for} \quad \omega \le 0).
\end{eqnarray}
where
\begin{eqnarray}
\gamma_{-,\omega}&=&\int dk |g_{\omega}(k)|^2\frac{-i(N(k)+1)}{\omega-\omega(k)-i0}\nonumber\\
&=&\pi \int dk~|g_{\omega}(k)|^2\frac{e^{\beta(\omega(k))\omega(k)}}{e^{\beta(\omega(k)\omega(k))}-1}\delta(\omega(k)-\omega)-i{\rm P.P}\int dk~
\frac{|g_{\omega}(k)|^2}{\omega(k)-\omega}\frac{e^{\beta(\omega(k))\omega(k)}}{e^{\beta(\omega(k))\omega(k)}-1}
\\
\gamma_{+,\omega}&=&\int dk |g_{\omega}(k)|^2\frac{-iN(k)}{\omega-\omega(k)-i0}\nonumber\\
&=&\pi \int dk~|g_{\omega}(k)|^2\frac{1}{e^{\beta(\omega(k))\omega(k)}-1}\delta(\omega(k)-\omega)-i{\rm P.P}\int dk~
\frac{|g_{\omega}(k)|^2}{\omega(k)-\omega}\frac{1}{e^{\beta(\omega(k))\omega(k)}-1}.
\end{eqnarray}
\end{subequations}
As in the previous model, the off-diagonal elements vanish when $G_{mn}\neq 0, ~(\forall~ m,n)$ which is defined in (\ref{eq:def_G_MN}).
In order to see if the stationary state can violate the detailed balance 
condition or not, let us check condition (\ref{db_condition}).
With direct computation we find
\begin{subequations}\label{quotient3}
\begin{equation}
\frac{\Gamma_{ml}}{\Gamma_{lm}}=e^{+\beta(\epsilon_{m}-\epsilon_{l})
(\epsilon_{m}-\epsilon_l)}\quad \mbox{for}\quad \epsilon_m >\epsilon_l
\end{equation}
\begin{equation}
\frac{\Gamma_{ml}}{\Gamma_{lm}}=e^{-\beta(\epsilon_{l}-\epsilon_{m})
(\epsilon_{l}-\epsilon_m)}\quad \mbox{for}\quad \epsilon_m <\epsilon_l
\end{equation}
\end{subequations}
Let us remark this fraction does not depend on the structure function
$g(k)$ unlike the previous model, however it can violate condition
(\ref{db_condition}) due to the generalized temperature function 
$\beta(\omega)$, i.e.
\begin{equation}
\frac{\Gamma_{ml}}{\Gamma_{lm}}
\neq
\frac{\Gamma_{mk}}{\Gamma_{km}}
\frac{\Gamma_{kl}}{\Gamma_{lk}}
\end{equation}
except for the constant temperature case (\ref{eq:constant_temperature}).

Let us show a typical example of non-equilibrium effects due to
the generalized temperature function. To realize the stationary state 
with non-detailed balance condition at least, two Bohr frequencies
(three level system) are necessary. With a generic 3--level system, whose energy 
levels are given by $\epsilon_1 < \epsilon_2 <\epsilon_3$ and $\epsilon_3-\epsilon_2\neq \epsilon_2-\epsilon_1$: 
the concrete form of the matrix $A$ in (\ref{eq:diagonal_equation}) 
is written as
\begin{equation}\label{matrix-A_3level}
A=\left(
\begin{array}{ccc}
\Gamma_{+,\epsilon_2-\epsilon_1}+\Gamma_{+,\epsilon_3-\epsilon_1}&
-\Gamma_{-,\epsilon_2-\epsilon_1}&-\Gamma_{-,\epsilon_3-\epsilon_1}\\
-\Gamma_{+,\epsilon_2-\epsilon_1}&
\Gamma_{-,\epsilon_2-\epsilon_1}+\Gamma_{+,\epsilon_3-\epsilon_1}&
-\Gamma_{-,\epsilon_3-\epsilon_2}\\
-\Gamma_{+,\epsilon_3-\epsilon_1}&
-\Gamma_{+,\epsilon_3-\epsilon_2}&
\Gamma_{-,\epsilon_2-\epsilon_1}+\Gamma_{-,\epsilon_3-\epsilon_2}
\end{array}
\right)
\end{equation}
and one can directly see that its eigenvalues are
\begin{subequations}
\begin{equation}
\lambda=0,\frac{b\pm\sqrt{b^2-4c}}{2}>0,
\end{equation}
\begin{eqnarray}
b&=&\Gamma_{+,\epsilon_2-\epsilon_1}+\Gamma_{+,\epsilon_3-\epsilon_1}+\Gamma_{+,\epsilon_3-\epsilon_2}+\Gamma_{-,\epsilon_2-\epsilon_1}+\Gamma_{-,\epsilon_3-\epsilon_1}+\Gamma_{-,\epsilon_3-\epsilon_2}\\
c&=&\Gamma_{+,\epsilon_2-\epsilon_1}\Gamma_{+,\epsilon_3-\epsilon_2}
+\Gamma_{+,\epsilon_2-\epsilon_1}\Gamma_{-,\epsilon_3-\epsilon_1}
+\Gamma_{+,\epsilon_2-\epsilon_1}\Gamma_{-,\epsilon_3-\epsilon_2}
+\Gamma_{+,\epsilon_3-\epsilon_1}\Gamma_{-,\epsilon_2-\epsilon_1}
+\Gamma_{+,\epsilon_3-\epsilon_1}\Gamma_{+,\epsilon_3-\epsilon_2}\nonumber\\
&&
+\Gamma_{+,\epsilon_3-\epsilon_1}\Gamma_{-,\epsilon_3-\epsilon_2}
+\Gamma_{-,\epsilon_2-\epsilon_1}\Gamma_{-,\epsilon_3-\epsilon_1}
+\Gamma_{-,\epsilon_2-\epsilon_1}\Gamma_{-,\epsilon_3-\epsilon_2}
+\Gamma_{+,\epsilon_3-\epsilon_2}\Gamma_{-,\epsilon_3-\epsilon_1}
\end{eqnarray}
\end{subequations}
and the stationary state
\begin{subequations}\label{st_sol}
\begin{equation}
\rho_{11}={1\over1+X+Y}\,;\ \rho_{22}={X\over1+X+Y}\,;\ \rho_{33}
={Y\over1+X+Y}
\end{equation}
where
\begin{eqnarray}\label{st_sol_2}
\frac{\rho_{22}}{\rho_{11}}&=&
\frac{\Gamma_{-,\epsilon_3-\epsilon_1}\Gamma_{+,\epsilon_2-\epsilon_1}+
\Gamma_{-,\epsilon_3-\epsilon_2}\Gamma_{+,\epsilon_2-\epsilon_1}+
\Gamma_{+,\epsilon_3-\epsilon_1}\Gamma_{-,\epsilon_3-\epsilon_2}}
{\Gamma_{-,\epsilon_3-\epsilon_1}\Gamma_{-,\epsilon_2-\epsilon_1}+
\Gamma_{-,\epsilon_3-\epsilon_1}\Gamma_{+,\epsilon_3-\epsilon_2}+
\Gamma_{-,\epsilon_3-\epsilon_2}\Gamma_{-,\epsilon_2-\epsilon_1}}\,=:X
\\
\frac{\rho_{33}}{\rho_{11}}&=&
\frac{
\Gamma_{+,\epsilon_3-\epsilon_1}\Gamma_{-,\epsilon_2-\epsilon_1}+
\Gamma_{+,\epsilon_3-\epsilon_2}\Gamma_{+,\epsilon_2-\epsilon_1}+
\Gamma_{+,\epsilon_3-\epsilon_1}\Gamma_{+,\epsilon_3-\epsilon_2}}
{\Gamma_{-,\epsilon_3-\epsilon_2}\Gamma_{-,\epsilon_2-\epsilon_1}+
\Gamma_{-,\epsilon_3-\epsilon_1}\Gamma_{-,\epsilon_2-\epsilon_1}+
\Gamma_{-,\epsilon_3-\epsilon_1}\Gamma_{+,\epsilon_3-\epsilon_2}}\,=:Y
\\
\frac{\rho_{33}}{\rho_{22}}&=&
\frac
{\Gamma_{+,\epsilon_3-\epsilon_2}\Gamma_{+,\epsilon_2-\epsilon_1}+
\Gamma_{+,\epsilon_3-\epsilon_1}\Gamma_{-,\epsilon_2-\epsilon_1}+
\Gamma_{+,\epsilon_3-\epsilon_2}\Gamma_{+,\epsilon_3-\epsilon_1}}
{\Gamma_{-,\epsilon_3-\epsilon_1}\Gamma_{+,\epsilon_2-\epsilon_1}+
\Gamma_{-,\epsilon_3-\epsilon_2}\Gamma_{+,\epsilon_2-\epsilon_1}+
\Gamma_{+,\epsilon_3-\epsilon_1}\Gamma_{-,\epsilon_3-\epsilon_2}}\,=:Z.
\end{eqnarray}
\end{subequations}
When (\ref{db_condition}) is not satisfied, the above solution
does not satisfy the detailed balance condition.
Notice that in this case the detailed balance condition is equivalent to
\begin{equation}\label{eq:insomesense_linear}
\delta:=\beta(\epsilon_2-\epsilon_1)(\epsilon_2-\epsilon_1)-
\beta(\epsilon_3-\epsilon_1)(\epsilon_3-\epsilon_1)
+\beta(\epsilon_3-\epsilon_2)(\epsilon_3-\epsilon_2)=0.
\end{equation}
Let us remark that the physics of this model can be different from
the previous model. For example,
taking $\langle \epsilon_1|D|\epsilon_2 \rangle=0$ 
(so as $\Gamma_{\pm,\epsilon_2-\epsilon_1}=0$) for simplicity, 
the above quotients become
\begin{subequations}
\begin{equation}
\frac{\rho_{22}}{\rho_{11}}=\frac{\Gamma_{+,\epsilon_3-\epsilon_1}\Gamma_{-,\epsilon_3-\epsilon_2}}{\Gamma_{-,\epsilon_3-\epsilon_1}\Gamma_{+,\epsilon_3-\epsilon_2}}=e^{\beta(\epsilon_3-\epsilon_2)(\epsilon_3-\epsilon_2)}
e^{-\beta(\epsilon_3-\epsilon_1)(\epsilon_3-\epsilon_1)}:=X
\end{equation}
\begin{equation}
\frac{\rho_{33}}{\rho_{11}}
=\frac{\Gamma_{+,\epsilon_3-\epsilon_1}}{\Gamma_{-,\epsilon_3-\epsilon_1}}
=e^{-\beta(\epsilon_3-\epsilon_1)(\epsilon_3-\epsilon_1)}<1,
\quad \frac{\rho_{33}}{\rho_{22}}
=\frac{\Gamma_{+,\epsilon_3-\epsilon_2}}{\Gamma_{-,\epsilon_3-\epsilon_2}}
=e^{-\beta(\epsilon_3-\epsilon_2)(\epsilon_3-\epsilon_2)}<1
\end{equation}
\end{subequations}
and $X$ is larger than $1$ when
\begin{equation}
\beta(\epsilon_3-\epsilon_2)
>
\frac{\epsilon_2-\epsilon_1}{\epsilon_3-\epsilon_2}
\beta(\epsilon_3-\epsilon_1)
\end{equation}
Thus, for such temperature function $\beta(x)$ the stationary state
satisfies $\rho_{22}>\rho_{11}$ which means that $2$ is a so-called
inversely populated state. 

Here, we focus on the current passing through the stationary state
and discuss the non-linear effects. 
For simplicity, we discuss the case of a three level system.
In this case ($\epsilon_1<\epsilon_2<\epsilon_3$), with direct computation
we obtain
\begin{subequations}\label{st_3_currentB}
\begin{eqnarray}
J_{mn}:&=&\int_{\Omega_{mn}} dk \langle U_t^\dagger n_k U_t\rangle
\nonumber\\
&=&(-1)^{m+n+1}\frac{e^{\beta(\epsilon_2-\epsilon_1)
(\epsilon_2-\epsilon_1)-\beta(\epsilon_3-\epsilon_1)(\epsilon_3-\epsilon_1)
+\beta(\epsilon_3-\epsilon_2)(\epsilon_3-\epsilon_2)}-1}
{(e^{\beta(\epsilon_2-\epsilon_1)(\epsilon_2-\epsilon_1)}-1)
(e^{\beta(\epsilon_3-\epsilon_2)(\epsilon_3-\epsilon_2)}-1)
(1-e^{-\beta(\epsilon_3-\epsilon_1)(\epsilon_3-\epsilon_1)})}I
\\
J^E_{mn}:&=&\int_{\Omega_{mn}} dk \omega(k)\langle U_t^\dagger n_k U_t\rangle
=(\epsilon_m-\epsilon_n)J_{mn}\\
I&=& |\langle\epsilon_1|D|\epsilon_2\rangle|^2~\int_{k\in \Omega_{21}} dk~|g(k)|^2\delta(\omega(k)-(\epsilon_2-\epsilon_1))\nonumber\\
&& \hspace*{1cm}\times~|\langle\epsilon_2|D|\epsilon_3\rangle|^2~\int_{k\in \Omega_{32}} dk~|g(k)|^2\delta(\omega(k)-(\epsilon_3-\epsilon_2))\nonumber\\
&& \hspace*{2cm}\times~|\langle\epsilon_3|D|\epsilon_1\rangle|^2~\int_{k\in \Omega_{31}} dk~|g(k)|^2\delta(\omega(k)-(\epsilon_3-\epsilon_1))
\end{eqnarray}
\end{subequations}
Notice $J^{(E)}_{21}$ and $J^{(E)}_{32}$ have same (and $J^{(E)}_{31}$
has opposite) sign. In addition
\begin{equation}\label{current3}
J^{(E)}_{31}=-(J^{(E)}_{21}+J^{(E)}_{32})
\end{equation}
and the sign of each currents depends on
\begin{equation}
\delta:=\beta(\epsilon_2-\epsilon_1)(\epsilon_2-\epsilon_1)
-\beta(\epsilon_3-\epsilon_1)(\epsilon_3-\epsilon_1)
+\beta(\epsilon_3-\epsilon_2)(\epsilon_3-\epsilon_2).
\end{equation}
In the case $\delta=0$, all currents vanish.
Especially, when the function is a constant $\beta$ 
(i.e. the initial state of the field is an equilibrium state with
temperature $\beta^{-1}$), this is easily understood with the fact
that the state of the system converges to the equilibrium state at the same
temperature without any stationary currents. Notice that even within the
linear approximation up to order $\delta$, there is no local
stationary state (with currents) which satisfies the detailed balance
condition, unlike the previous model. In this model, the existence of currents 
always implies the deviation from the equilibrium.

Now let us see some interesting properties of the currents
(\ref{st_3_currentB}). 
In the case $\delta >0$, we obtain from 
(\ref{st_3_currentB}) 
\begin{equation}\label{down_conver}
J^{(E)}_{21}, J^{(E)}_{32} >0,\quad \mbox{and}\quad J^{(E)}_{13}<0. 
\end{equation}
As clearly understood from the definition of the currents,
the relation (\ref{down_conver}) is describing the process that a field
quantum with energy $\epsilon_3-\epsilon_1$ is converted into two quanta with energy $\epsilon_2-\epsilon_1$ and $\epsilon_3-\epsilon_2$.
On the contrary when $\delta <0$,
\begin{equation}
J^{(E)}_{21}, J^{(E)}_{32} <0,\quad \mbox{and}\quad J^{(E)}_{13}>0 
\end{equation}
and this can be interpreted as a process from two quanta to one quantum.
There are interesting analogies of these processes
with parametric downconversion and second harmonic generation in non-linear
quantum optics\cite{Walls95}.
They are considered as opposite process of another.
In our model, the direction of the process depends on the generalized
temperature function $\beta(\omega)$ which is a parameter of the
initial state of the field.
This phenomenon can be understood as the fact that through interaction
with a non-equilibrium field the system can have such a function,
which is an example of dissipative structure in the Prigogine sense
\cite{Prigogine62}.

\section{Discussion}\label{sec:sum}
In conclusion, let us  further comment on a few related topics.

\begin{description}
\item[1)]
On the irreversibility and unitarity of time evolution.

As we discussed in Sec.\ref{subSec.linear_app},
we can see irreversibility in this model through the entropy production
(\ref{entropy_production}) due to the stationary currents, which 
should be considered as processes involving the total system including
the environment. 
On the other hand, the time evolution operator of the total system
$U_t$ is unitary in the sense that
\begin{equation}\label{unitarity condition}
U^\dagger_t U_t= U_t U_t^\dagger=1, \quad t>0,
\end{equation}
which is easily checked by putting $X=1$ in (\ref{eq:lange}).
These statements might seem to be contradiction.
However, one should notice that the appearance of irreversibility
has nothing to do with the unitarity of $U_t$.
When the temperatures of both environments are the same,
it is known that the unitarity condition (\ref{unitarity condition}) is
required to realize a physical fluctuation-dissipation relation or
a correct equilibrium stationary state \cite{AcLuVo}.
Moreover, when we speak of macroscopic phenomena like entropy production,
we need a good procedure to extract the proper degrees of
freedom to discuss them.
Since there exist same macroscopic states which are distinguishable
microscopically from each other, not all microscopic degrees can be employed
to discuss macroscopic properties.
Indeed, the entropy production (\ref{entropy_production}) is discussed
in terms of what we call slow degrees of freedom,
and the stochastic limit can be considered as the procedure of extracting
the proper degrees of freedom. 
In other words, we extract information from the total dynamics
as slow degrees which can describe the macroscopic phenomena.

\item[2.)] Local KMS condition for field.

A possible formulation of the local KMS condition for the field is the
following.

\begin{definition}\label{lockms} A state $\langle\cdot\rangle$ on the
polynomial algebra $a_k$, $a^\dagger_k$, is said to satisfy the local KMS
condition with temperature function $\beta:\mathbb R^d\to\mathbb R$ if, for
every $m$, $n\in\mathbb N$, $\varepsilon_1,\dots,\varepsilon_n$,
$\eta_1,\dots,\eta_m\in\{0,1\}$, and with the convention $x^0=x^\dagger$,
$x^1=x$ for any operator $x$, the following identities hold in the sense
of distributions.
\end{definition}

$$\langle a^{\eta_1}_{k_1}(0)\dots a^{\eta_m}_{k_m}(0)a^{\var_n}_{h_n}
(t+i\beta_{h_n})a^{\var_{n-1}}_{h_{n-1}}(t+i\beta_{h_{n-1}})
\dots a^{\var_1}_{h_1}(t+i\beta_{h_1})\rangle$$
\begin{equation}
=\langle a^{\var_n}_{h_n}(t)\dots a^{\var_1}_{h_1}(t)a^{\eta_1}_{k_1}
(0)\dots a^{\eta_m}_{k_m}(0)\rangle.
\label{lkms}
\end{equation}

\begin{lemma} Define the local inverse temperature function by
\begin{equation}
-\beta(k):=\left(\log{n(k)\over m(k)}\right){1\over\omega_k},
\label{loctemp}
\end{equation}
where
\begin{equation}
\langle a_ka^\dagger_{k'}\rangle =: m(k)\delta(k-k') =
\frac{e^{\beta\omega_k}}{e^{\beta\omega_k}-q}=(qn(k)+1)\delta(k-k')
\label{qgibfa2}
\end{equation}
($q=-1$ for Bosons and $q=+1$ for Fermions).
Then the local KMS condition is satisfied by the 2--point functions:
\begin{equation}
\langle a_k(0)a^\dagger_{k'}(t+i\beta(k'))\rangle=\langle
a^\dagger_{k'}(t)a_k\rangle\label{aa+}
\end{equation}
\begin{equation}
\langle a^\dagger_k(0)a_{k'}(t+i\beta(k'))\rangle=\langle
a_{k'}(t)a^\dagger_k(0)\rangle\label{a+a}
\end{equation}
\end{lemma}

\noindent{\it Proof\/}. In the above notations, one has
$$\langle a_k(0)a^\dagger_{k'}(t+i\beta(k'))\rangle=e^{i(t+i\beta(k'))\omega_{k'}}
\langle a_ka^\dagger_{k'}\rangle=e^{-\beta(k')\omega_{k'}}e^{it\omega_{k'}}\langle
a_ka^\dagger_{k'}\rangle$$
$$=e^{it\omega_k}{m(k)\over n(k)}\,n(k)\delta(k-k')
=e^{it\omega_{k'}}m(k')\delta(k'-k)=e^{it\omega_{k'}}\langle
a^\dagger_{k'}a_{k}\rangle=\langle a^\dagger_{k'}(t)a_k\rangle$$
and this proves (\ref{aa+}). In a similar way one verifies that
(\ref{a+a}) holds.

\begin{proposition}\label{lkmsgau} If the state $\langle\cdot\rangle$ is
mean zero gauge invariant and Boson Gaussian then condition (\ref{lkms}) is
satisfied.
\end{proposition}

\noindent{\it Proof\/}. By Gaussianity both sides of (\ref{lkms}) are
reduced to weighted sums of pair correlation functions. Since in both
sides of (\ref{lkms}) we can distinguish the $(h,\varepsilon)$--terms
from the $(k,\eta)$--terms and since the pair correlations preserve the
order, there will be 3 types of pair correlations:
(i) those of type $(h,k)$, (ii) those of type $(h,h)$, 
(iii) those of type $(k,k)$.

In case (i), due to gauge invariance, the only none zero combinations are
of the form $\langle aa^\dagger\rangle$ or $\langle a^\dagger a\rangle$ so we can
apply (\ref{aa+}) and (\ref{a+a}).

In case (ii) the terms are already in the correct order.

In case (iii), again by gauge invariance, the only possibilities are
$$\langle a_h(t+i\beta_h)a^\dagger_{h'}(t+i\beta_{h'})\rangle=
e^{-(t+i\beta_h)\omega_h}e^{i(t+i\beta_{h'})\omega_{h'}}
\langle a_ha^\dagger_{h'}\rangle$$
$$=e^{it(\omega_{h'}-\omega_h)+(\beta_h\omega_h-\beta_{h'}\omega_{h'})}
\delta(h-h')$$
\begin{equation}
=\langle a^\dagger_h(t)a_{h'}(t)\rangle\label{kktrm}
\end{equation}
and similarly for the other term.

Since in the Boson case the weight of each pair partition is equal to 1,
after the replacements (\ref{aa+}), (\ref{a+a}), (\ref{kktrm}) the
pair--partition expansion of the left hand side of (\ref{lkms}) becomes
the pair--partition expansion of the right hand side.\bigskip

The validity of the local KMS condition for more
general Gaussian states as well as for quantum Markov states is now
under investigation.

\item[3.)] The generalized temperature function and its thermodynamics.

On the description of the generalized temperature function $\beta(H)$,
R. S. Ingarden, A. Kossakowski, M. Ohya, T. Nakagomi
have discussed similar idea
in the context of information theory~\cite{Ingarden}.
They introduce a system described by the density operator
\begin{equation}
\rho=\frac{1}{Z(\beta_1,...,\beta_n)}\exp
\left(-\sum_{j=1}^{n}\beta_{j}H^j\right),\quad \beta_j >0
\end{equation}
and  discussed possible generalization of thermodynamics
for structured complex systems (e.g. biological system)
including bifurcations, catastrophes and self organization.
As mentioned in their book \cite{Ingarden}, their phenomenological
idea is in the line of thought of synergetics by Haken~\cite{Haken}.
In the present paper, we explained the microscopic origin of such states
and their physical meaning through the dynamical detailed balance condition.
Through the local KMS condition
a general classification of such non-equilibrium states became possible.
We believe that our approach gives a good insight to
generalization of thermodynamics in this direction.
\end{description}


\begin{thebibliography}{}
\bibitem{Prigogine62}
I. Prigogine, {\it Nonequilibrium statistical mechanics},
New York, Wiley(1962);
P. Glansdorff, I.Prigogine, {\it Thermodynamic theory of structure, stability 
and fluctuations},  Wiley-Interscience, London, (1971).

\bibitem{Zubarev}
D. N. Zubarev, {\it Nonequilibrium Statistical Thermodynamics},
Consultants, New York, (1974).

\bibitem{Toda}
M. Toda, R. Kubo and N. Saito,
{\it Statistical Physics I}, Springer New York, (1992);
R. Kubo, M. Toda, and N. Hashizume,
{\it Statistical Physics II}, Springer New York (1997).

\bibitem{Spohn-Lebowits78}
H. Spohn, J.~L. Lebowitz, Commun math phys. {\bf 54}, (1977) 97;
Adv. Chem. Phys. {\bf 38}, 109 (1978) and
reference therein.

\bibitem{Antoniou}
I. Antoniou and S. Tasaki, Int. J. Quantum Chemistry {\bf 46},(1993),
425-474;
I. Antoniou, K. Gustafson, Physica A {\bf 236},(1997), 296-308

\bibitem{Tasaki}
S. Tasaki, Quantum information, III (Nagoya, 2000), 157,
World Sci. Publishing, River Edge, NJ, 2001;
Chaos Solitons Fractals 12 (2001), no. 14-15, 2657.
and reference therein.

\bibitem{Jalsic02}
V. Jak\v{s}i\'{c}, C.-A. Pillet, Commun. Math. Phys. {\bf 226}
(2002), 131-162 and reference therein.

\bibitem{Schmuser02}
F. Schmuser, B. Schmittmann,
J. Phys. A {\bf 35} (2002), 2567.

\bibitem{Bedeaux01}
D. Bedeaux, P. Mazur, Phys. A 298 (2001), 81-100

\bibitem{Ojima}
D. Buchholz, I. Ojima, H. Roos,
Annals Phys. {\bf 297} (2002) 219-242

\bibitem{stochastic_limit}
L. Accardi, A. Frigerio, and Y.~G. Lu,
Commun. Math. Phys. {\bf 131}, 537 (1990);
L. Accardi, J. Gough, and Y.~G. Lu,
Rep. Math. Phys. {\bf 36}, 155 (1995);
L. Accardi, S.~V. Kozyrev, and I.~V. Volovich,
Phys. Lett. A {\bf 260}, 31 (1999); G. Kimura, K. Yuasa, and
K. Imafuku, Phys. Rev. A {\bf 63} (2001), 022103;
Phys. Rev. Lett. (2002), in printing.

\bibitem{AcLuVo}
L. Accardi, Y. G. Lu, and I. V. Volovich,
\textit{Quantum Theory and Its Stochastic Limit}
Springer-Verlag (2002).

\bibitem{Accardi-Fagnola}
L. Accardi, F. Fagnola (eds.), ~{\it Quantum interacting particle systems}, 
Lecture Note of Levico school, September 2000, Volterra Preprint N.431.
in {\it Quantum interaction particles} L. Accardi, F. Fagnola (eds.)
World Scientific (2002).

\bibitem{Accardi-Imafuku-Kozyrev02}
L. Accardi, K. Imafuku, S. V. Kozyrev, {\it
Proc. of XXII Solvay Conference in Physics, 24-29 November 2001}, 
Springer, (2002).

\bibitem{Accardi-Imafuku-Lu02}
L. Accardi, K. Imafuku, Y.~G. Lu, in preparation


\bibitem{Gorini-Kossokowski76}
V. Gorini, V. Kossakowski, J. Math. Phys. {\bf 17}, 1298 (1976);
J. Math. Phys. {\bf 17}, 2123 (1976);
R. Alicki, Rep. Math. Phys. {\bf 10}, 249, (1976)

\bibitem{Kossakowski-Frigerio-Gorini-Verri77}
A. Kossakowski, A. Frigerio, V. Gorini, and M. Verri, Commun. Math. Phys.
{\bf 57}, 97, (1977); 
A. Frigerio and V. Gorini, Commun. Math. Phys. {\bf 93}, 517 (1984)
and reference therein.

\bibitem{Onsager31}
L. Onsager, Phys. Rev. {\bf 37}, 405 (1931); Phys. Rev. {\bf 38}, 2265 (1931).

\bibitem{GKSL}
V. Gorini, A. Kossakowski, and E. C. G. Sudarshan, 
J.Math. Phys. {\bf 17}, 821 (1976); V. Gorini et al.,
Rep. Math. Phys. {\bf 13}, 149 (1978).
G. Lindblad, Commun. Math. Phys. 48, 119 (1976).


\bibitem{Kubo57}
R. Kubo, M. Yokota and S. Nakajima, J. Phys. Soc. Jpn. {\bf 12} 1203
(1957).

\bibitem{Jona96}
D. Gabrielli, G. Jona-Lasinio and C. Landim,
Phys. Rev. Lett. {\bf 77}, 1202 (1996).


\bibitem{Walls95}
For example, D. F. Walls, G. J. Milburn, {\it Quantum Optics},
Springer-Verlag Berlin and Heidelberg (1995).

\bibitem{Ingarden}
R. S. Ingarden, Bull. Acad. Polon. Sci., S\'{e}r. math. astr. phys. {\bf 11} (1963); R. S. Ingarden and A. Kosakowski, Bull. Acad. Polon. Sci., S\'{e}r. math. astr. phys. {\bf 16} (1968); T. Nakagomi, Open Sys. Information Dyn. {\bf 1},
233 (1992); R.S. Ingarden, A. Kossakowski, and M. Ohya,
``Information Dynamics and Open systems"
Kluwer Accademic Publishers (1997), and reference therein.

\bibitem{Haken}
H. Haken, ``Synergetics, An Introduction", Springer, Berlin (1983),
``Advanced Synergetics", Springer, Berlin (1987),
``Information and Self-Organization", Springer, Berlin (1988).

\end{thebibliography}
\end{document}